\documentclass{aa}  
\usepackage{txfonts}
\usepackage{graphicx}	
\usepackage{amsmath}	
\usepackage{bm}
\usepackage{natbib}
\bibpunct{(}{)}{;}{a}{}{,}
\usepackage{hyperref}

\begin{document}

   \title{Dynamical constraints on the S2 (S0-2) star possible companions} 

   \author{Rodrigo P. Silva \inst{1,2,3}
          \and
          Alexandre C. M. Correia\inst{1,4}
          \and
          Tjarda C. N. Boekholt \inst{5}
          \and
          Paulo J. V. Garcia \inst{2,3}
          }

   \institute{
        CFisUC, Departamento de F\'isica, Universidade de Coimbra, 3004-516 Coimbra, Portugal 
        \and
        Faculdade de Engenharia, Universidade do Porto, Rua Dr. Roberto Frias, 4200-465 Porto, Portugal
        \and
        CENTRA - Centro de Astrof\'isica e Gravita\c c\~ao, IST, Universidade de Lisboa, 1049-001 Lisboa, Portugal
        \and
        LTE, Observatoire de Paris, Universit\'e PSL, Sorbonne Universit\'e, CNRS, 75014 Paris, France
        \and
        Anton Pannekoek Institute for Astronomy, University of Amsterdam, NL-1090 GE Amsterdam, The Netherlands
        }

\date{\today; Received; accepted To be inserted later}

\abstract{
The center of the Galaxy harbors a supermassive black hole, Sgr\,A*, which is surrounded by a massive star cluster known as the S-cluster. The most extensively studied star in this cluster is the B-type main-sequence S2 star (also known as S0-2). These types of stars are commonly found in binary systems in the Galactic field, but observations do not seem to detect a companion to S2. This absence may be attributed to observational biases or to a dynamically hostile environment caused by phenomena such as tidal disruption or mergers. Using a $N$-body code with first-order post-Newtonian corrections, we investigate whether S2 can host a stellar or planetary companion. We perform $10^{5}$ simulations adopting uniform distributions for the orbital elements of the companion.
Our results show that companions may exist for orbital periods shorter than 100~days, eccentricities below 0.8, and across the full range of mutual inclinations. The number of surviving companions increases with shorter orbital periods, lower eccentricities, and nearly coplanar orbits. We also find that the disruption mechanisms include mergers driven by Lidov-Kozai cycles and breakups that occur when the companion surpasses the Hill radius of its orbit. Finally, we find that the presence of a companion would alter S2's astrometric signal by no more than $ 5\,\mu{\rm as}$. Current radial-velocity detection limits constrain viable stellar binary configurations to approximately 4.4\% of the simulated cases. Including astrometric limits reduces to 4.3\%. Imposing an additional constraint that any companion must have a mass $\lesssim 2\,M_{\odot}$ (otherwise it would be visible) narrows the fraction of undetectable stellar binaries to just 3.0\%.
}


   \maketitle

\section{Introduction}

The S2-star is first referred to in \citet{1997MNRAS.284..576E} as a high proper-motion star, later found to be orbiting Sgr\,A* \citep{2002Natur.419..694S}, the supermassive black-hole (SMBH) at the Galactic center. 
It was spectroscopically detected by \citet{2003ApJ...586L.127G}, and deep spectroscopy by \citet{2008ApJ...672L.119M} showed that it is a young B0-B2.5V dwarf (main-sequence) star.
\citet{2017ApJ...847..120H} derived the mass, radius, and age of S2 using model atmospheres and stellar evolution models.

Precision radial velocity and especially astrometric measurements of S2 have: a) determined the distance to the center of the Milky Way with sub-percent uncertainty \citep{2019A&A...625L..10G}; b) measured Sgr\,A*  mass and tested General Relativity through gravitational redshift and Schwarzschild precession \citep{2018A&A...615L..15G, 2019Sci...365..664D, 2020A&A...636L...5G, 2021A&A...647A..59G}; c) derived upper limits to extended mass distributions \citep{2022A&A...657L..12G, 2024A&A...692A.242G}; d) tested the equivalence principle \citep{2019PhRvL.122j1102A}; e) investigated the existence of intermediate massive black holes in the vicinity of Sgr\,A* \citep{2023A&A...672A..63G}; f) probed for dark matter close to the supermassive black hole \citep[e.g.][]{2023PhRvD.108j1303D, 2023MNRAS.524.1075F, 2024MNRAS.527.3196S, 2024MNRAS.530.3740G}; and g) researched alternatives to General Relativity and new physics \citep[cf.][for a review]{2023RPPh...86j4901D}. The orbit of S2 is and will be monitored in the following decades with GRAVITY+ \citep{2022Msngr.189...17A} and the ELTs \citep{2017arXiv171106389D, 2024SPIE13096E..11S}.

S2 is a member of the S-star cluster dynamical group, the closest cluster to Sgr\,A* \citep[e.g.][]{2017ARA&A..55...17A}. The origin of this young cluster has remained elusive, which is not surprising given the central supermassive black hole gravitational influence \citep[e.g.][]{2005PhR...419...65A, 2010RvMP...82.3121G}. This cluster presents a reduced binarity fraction when compared with the young nuclear cluster in the central parsec \citep{2023ApJ...948...94C, 2024ApJ...964..164G} and similar type stars in the galaxy \citep[e.g.][]{2012Sci...337..444S,2013ARA&A..51..269D}. Key dynamical effects in the S-star cluster may contribute to it: a) individual stellar collisions \citep[e.g.][]{2023ApJ...955...30R}; b) resonant relaxation of the individual orbital orientation vectors \citep[e.g.][]{2015MNRAS.448.3265K}; c) stellar binary eccentricity and inclination oscillations such as the Lidov-Kozai mechanism \citep[e.g.][]{2016ARA&A..54..441N}; d) stellar binary evaporation \citep[e.g.][]{2016MNRAS.460.3494S}; e) stellar binary breakup by the Hills mechanism \citep[e.g.][]{2010ApJ...713...90A, 2024ApJ...977..268Y}; and f) star-star scattering \citep[e.g.][]{2019ApJ...875...42T, 2023MNRAS.526.5791P}. Their net effect is to reduce the binarity fraction with time, either via mergers or disruption. Interestingly, the population of G-clouds \citep[e.g.][]{2012Natur.481...51G} is thought to be created by mergers \citep{2020Natur.577..337C} and shares the same dependence of the pericenter distance to the orbital eccentricity as the S-stars \citep{2024ApJ...962...81B}.  

\citet{2017A&A...602A..94G} probed the binarity of S2, and found no companions up to 3~mag fainter in the K-band. This was expected since VLTI/GRAVITY cannot resolve the Hill sphere of the Sgr\,A* $-$ S2 system. Conversely, radial velocity series are more sensitive to inner orbits. The search by \citet{2018ApJ...854...12C} found no companions inside the Hill radius with $m_{\mathrm{B}} \sin I \gtrsim 2\,\mathrm{M}_\odot$ or 4~mag fainter.

Constraining the binarity nature of S2 holds clues on its dynamical past and is key for current and forthcoming high-order tests of gravity \citep[e.g.][]{2018MNRAS.476.3600W}. The individual binarity nature of the objects can also bias statistical properties \citep[e.g.][]{2018ApJ...853L..24N}. In this paper, we aim to address two key questions: a) Can S2 be a binary? and (b) If S2 is a binary, what is the resulting astrometric bias? To this end, we combine Monte Carlo dynamical simulations with observational constraints.

\section{Methods}\label{Methods}

\subsection{S2 companion statistical distribution} \label{sec:distributions}

\begin{table}
    \caption[]{Initial orbital elements and parameters of the Sgr\,A* $-$ S2 system used in the numerical simulations. \label{orb_elem_S2}}
        \begin{center}
        \begin{tabular}{|c|c|c|}
        \hline
        \textbf{element}  & \textbf{[unit]} & \textbf{value}  \\
        \hline
        $a_{\bullet}$& ["]  & 0.12497 \\
        $e_{\bullet}$  &              & 0.88441      \\
        $I_{\bullet}$ &[$^\circ$] & 134.69241  \\
        $M_{\bullet}$ &[$^\circ$] & 180.0  \\ 
        $\omega_{\bullet}$ & [$^\circ$] & 66.28411  \\
        $\Omega_{\bullet}$ & [$^\circ$] & 228.19245 \\   
        $m_{\bullet}$ &[$\mathrm{M}_\odot$] & $4.3\times10^{6}$ \\
        $m_{\mathrm{A}}$ &[$\mathrm{M}_\odot$] & 13.60 \\
        $R_{\mathrm{A}}$ &[$\mathrm{R}_\odot$] & 5.53 \\
        \hline
        \end{tabular}
        \end{center}
\end{table}

The orbital elements of S2 are taken from Table~1 in \citet{2023MNRAS.524.1075F}, with the exception of the mean anomaly, $M_{\bullet}$, which we set to $180.0^\circ$ to initiate the simulations at the point of minimal perturbation. The mass and radius of S2 are adopted from Table~2 in \citep{2017ApJ...847..120H}. All parameter values used in the simulations are summarized in Table~\ref{orb_elem_S2}.
The binary companion orbital elements are randomly generated. The mass ratio, $q = m_{\mathrm{B}}/m_{\mathrm{A}}$, where $m_{\mathrm{B}}$ is the mass of the companion, is uniformly sampled between 0.01 and 1.
The longitude of the node, $\Omega$, the argument of the pericenter $\omega$, and the mean anomaly, $M$, are uniformly sampled between 0 and $2 \pi$.
The orbital inclination, $I$, is generated by uniformly sampling $\cos I$ between $-1$ and $1$, to warrant a well-balanced distribution over the sphere.
For the orbital period, $P$, we uniformly sample $\log P$ for $P_\mathrm{R} \le P \le P_\mathrm{H}$, where $P_\mathrm{H} = 132.59~\mathrm{d}$ corresponds to the maximum orbital period allowed by the Hill sphere \citep[e.g.][]{1999ssd..book.....M}
\begin{equation}
r_\mathrm{H} 
= a_{\bullet}\left(1 - e_{\bullet} \right)\left(\frac{m_{\mathrm{A}}}{3m_{\bullet}} \right)^{1/3}  \ ,
\label{hill_limit}
\end{equation}
and $P_\mathrm{R} = 1.559~\mathrm{d}$ corresponds to the orbital period of a circular orbit at the Roche limit \citep{Roche_1849},
\begin{equation}
r_\mathrm{R} 
= f_{t} \, \left(1+q\right)^{1/3} R_{\mathrm{A}} \ ,
\label{roche_limit}
\end{equation}
with $f_{t} = 2.44$ \citep{Jeans_1919B}, and we assume for simplicity that all companions have the same mean density of S2. 
The eccentricity, $e$, is also uniformly sampled between $0$ and $1$. 
Still, we discard initial conditions for which the pericenter is below the Roche limit, i.e., $a \, (1-e) < r_\mathrm{R}$, where $a$ is the semi-major axis corresponding to $P$.
A reasonable balance is achieved between populating the sampled bins with a sufficient number of simulations and overall hypothesis simplicity.

\subsection{Orbital evolution}

We use the open-source $N$-body code \texttt{TIDYMESS} \citep{2023MNRAS.522.2885B}, which employs a time-symmetric adaptive timestep with post-Newtonian (PN) corrections, allowing accurate treatment of close encounters while conserving total energy \citep{2023MNRAS.519.3281B}. 
Many binary systems in our simulations become highly eccentric, chaotic, or disrupted, necessitating a variable timestep to accurately track their orbital evolution.

We assume point-mass bodies together with 1PN corrections for a Schwarzschild metric\footnote{Options \texttt{tidal\_model = 0}, \texttt{pn\_order = 1}, \texttt{dt\_mode = 2}, and \texttt{eta = 0.0625}.}.
Although \texttt{TIDYMESS} can account for tides in chaotic regimes, tidal effects are not considered because the minimum orbital period permitted by the Roche limit in our simulations is greater than the maximum period of tidal circularized B-type binaries \citep{2002ApJ...573..359A}.
Moreover, the general relativity precession timescale is much shorter than the tidal and rotational evolution timescales, and so these two effects can be neglected (see Appendix~\ref{Appendix:Distortions} for more details).
\texttt{TIDYMESS} thus reduces to a general point-particle, fourth-order $N$-body integrator, commonly used in stellar dynamics.

The S2 orbit always starts at the apocenter with the orbital elements previously indicated (Table~\ref{orb_elem_S2}).
We generate a sample of $10^5$ different initial hierarchical three-body systems (binary stars + SMBH) and evolve them for $10^6$~yr.
This time length value corresponds to a compromise between the computational running time and the dynamical evolution of the system.
The simulation is stopped whenever the binary system is disrupted.
This situation arises either because the distance between the two stars becomes lower than $r_\mathrm{R}$, corresponding to a merger, or because the orbital energy of the binary becomes positive, corresponding to a breakup event.

\section{Results}
\label{results}

\subsection{Disruption rate over time} 

\begin{figure}
    \centering
    \includegraphics[width=\columnwidth]{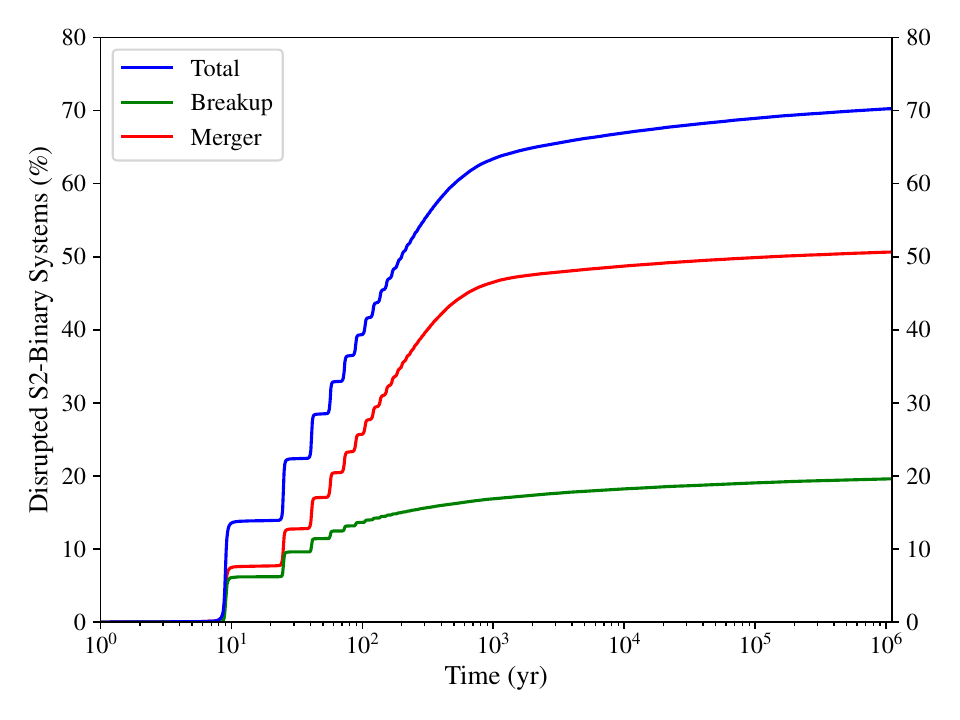}
    \caption{Disruption rate of binary systems over time. In red, we show the number of disrupted systems due to mergers; in green, those disrupted due to gravitational breakup; and in blue, the total number of disruptions.  \label{bin_frac}}
\end{figure}

\begin{figure*}
\centering
\includegraphics[width=0.9\columnwidth]{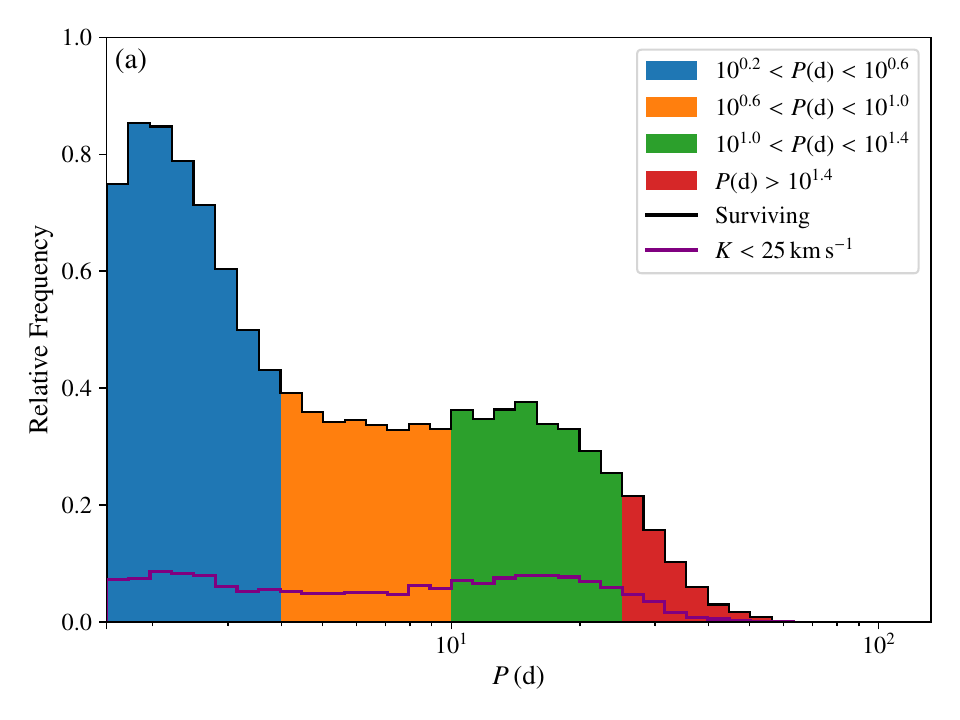} 
\includegraphics[width=0.9\columnwidth]{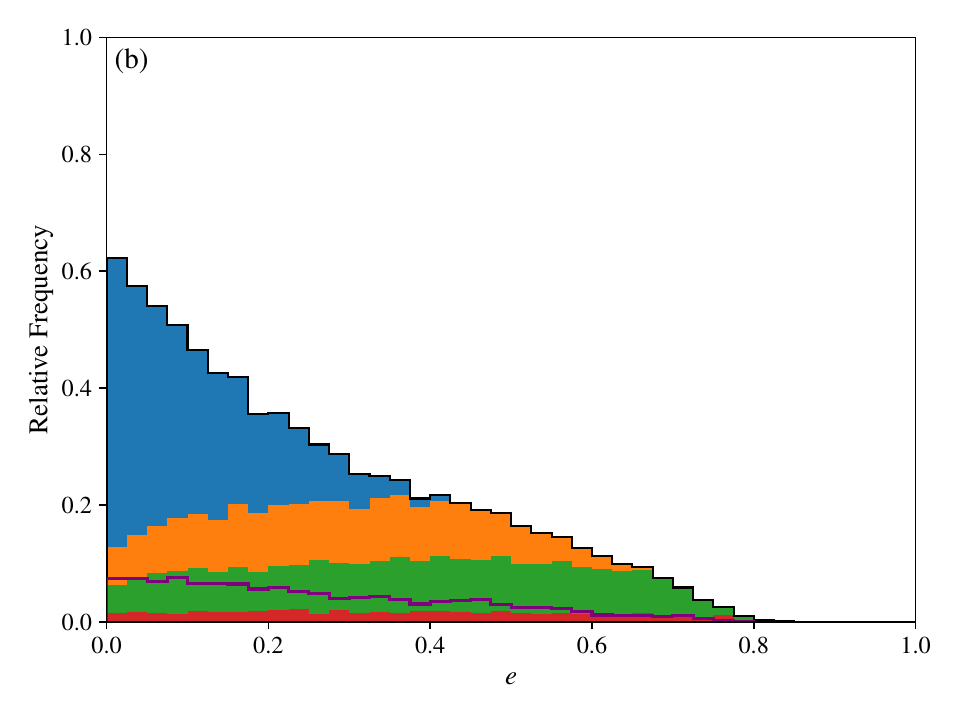} \\
\includegraphics[width=0.9\columnwidth]{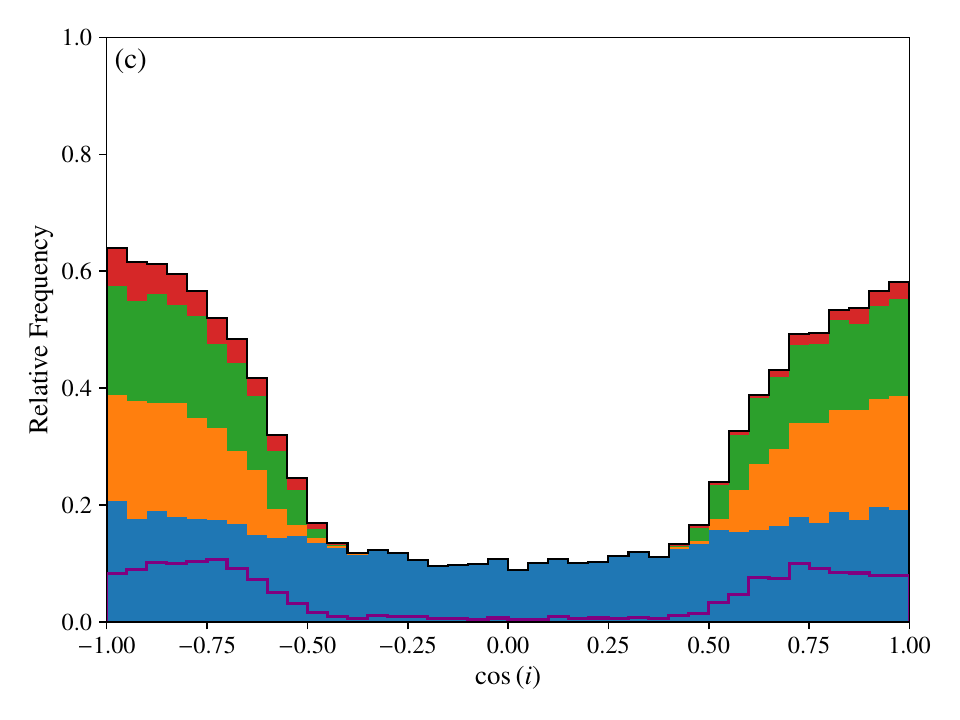} 
\includegraphics[width=0.9\columnwidth]{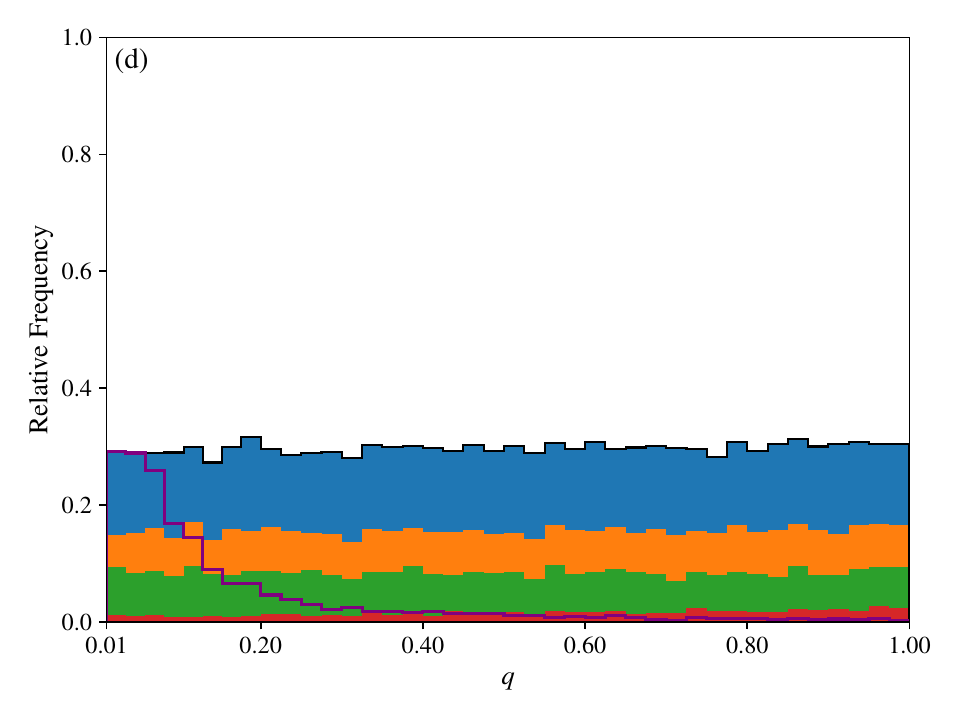}
\caption{Histograms of the relative frequency distribution of the surviving binary's main orbital parameters. We show the orbital period (a), 
the eccentricity (b), the mutual inclination (c), and the mass ratio (d). The black line gives the total distribution of the surviving systems, while the purple line represents the fraction that would have remained imperceptible to current radial-velocity measurements (Eq.\,(\ref{K-amplitude})). The color code corresponds to different ranges of orbital periods (a). \label{P_end}}
 \label{a:lambda}
\end{figure*}

In Fig.~\ref{bin_frac}, we show the temporal evolution of the disruption statistics. 
At the beginning of the simulations, we observe a significant increase in the number of mergers and breakups with cycles around every 16 years, which corresponds to the passage of S2 at the pericenter \citep{2017ApJ...837...30G}, where the gravitational disturbance of the SMBH is more substantial.

At the end of the simulations, most systems have been destroyed. Due to short-term perturbations, the destruction rate is very fast, with more than 60\% of the systems being destroyed in less than $10^3$~yr. After that time, the disruption rate slows down and can be mainly attributed to secular perturbations. About half of the disrupted systems correspond to a merger, while the other half are due to a breakup event.
The mergers are essentially driven by the Lidov-Kozai mechanism, while the breakups result from the binary surpassing the Hill radius of the orbit (see Appendix~\ref{Appendix:AnalyticalPredictions} for more details).

The number of surviving systems stabilizes around $10^6$~yr at only about 30\%. 
Inspecting the binary orbital parameters distributions of the systems at this plateau provides valuable insights into which configurations can survive near the SMBH.

\subsection{Orbital parameters of surviving systems}
\label{binaries}

In Fig.~\ref{P_end}, we present the relative frequency distribution of the orbital parameters of the surviving binaries. The frequency in each bin is computed by dividing the surviving systems by the initial number of systems.
In Fig.~\ref{P_end}\,(a), we show the distribution of orbital periods.
As expected, the number of surviving systems decreases with increasing orbital period, since wider binaries possess lower binding energies.
However, there are some unforeseen features.
Disruptions also occur for close-in orbits; for $P < 2.5$~d, only about 80\% survive. For orbital periods $2.5 < P < 16$~d, the number of survivors sharply decreases to about 40\%.
At this stage, there is a plateau in the number of survivals, and we can even spot a slight increase within $10 < P < 16$~d.
After that point, there is another step of decrease and no surviving systems are observed with orbital periods longer than 100~d.
To better understand the initial peak, the plateau, and the middle peak, we have attributed different colors to each region.
This color scheme is consistently used throughout the other panels in Fig.~\ref{P_end}, allowing to emphasize behaviors corresponding to each period segment.

\begin{figure}
    \centering
    \includegraphics[width=0.95\columnwidth]{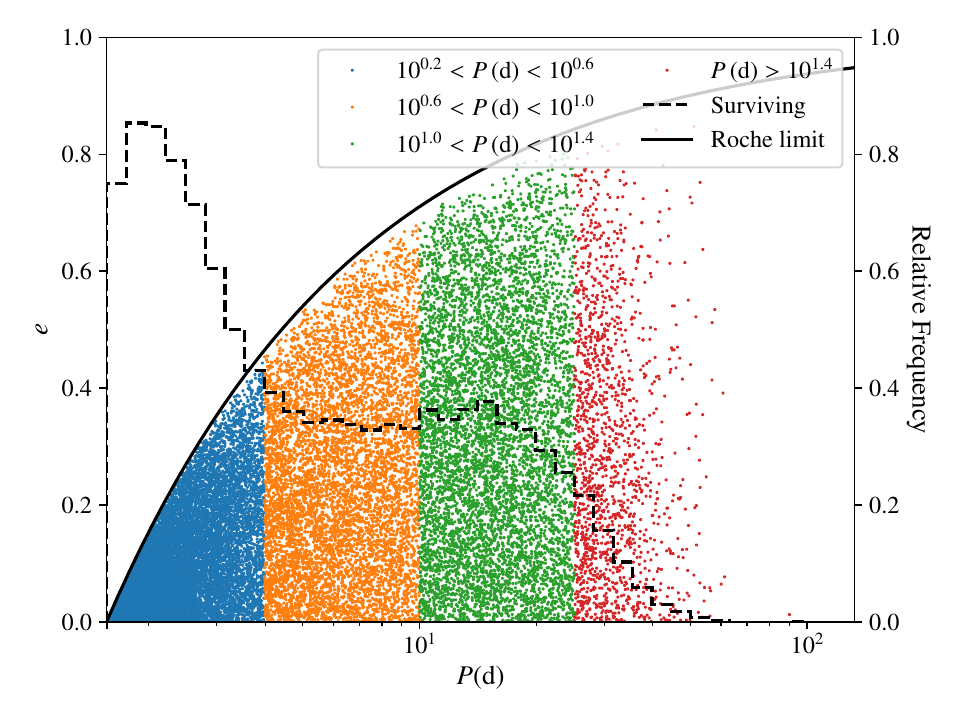}
    \caption{Distribution of the surviving binary systems as a function of the orbital period and eccentricity. The black curve gives the Roche limit (Eq.\,(\ref{roche_limit})). The color code corresponds to different ranges of orbital periods, and the black dashed line displays the histogram distribution of the surviving binaries (Fig.~\ref{P_end}\,(a)). \label{Roche_limit_Fig}}
\end{figure}

In Fig. \ref{P_end}\,(b), we show the distribution of the eccentricities.
As expected, the number of surviving systems decreases with the eccentricity because elliptical orbits experience a wider separation at the apocenter.
In particular, we do not observe any survival for $e>0.8$.
Elliptical orbits also lead to a shorter pericenter distance, facilitating approaches within the Roche limit that disrupt the system.
Indeed, we observe that close-in orbits (in blue) have a maximal eccentricity of about $0.4$, because in this period range higher eccentricities lead to a merger (see Fig.~\ref{Roche_limit_Fig}).
This feature also explains the plateau and the middle peak in the orbital period distribution (Fig.~\ref{P_end}\,(a)).
As the period increases, the systems become more susceptible to breakup events, but on the other hand, they become more resilient to Roche disruptions.
In the plateau region (in orange) the maximal eccentricity is about 0.67, while in the middle peak region (in green) the eccentricity is no longer a constraint (Fig.~\ref{Roche_limit_Fig}).

In Fig. \ref{P_end}\,(c), we show the distribution of the mutual inclinations.
We observe that many binary systems are disrupted for $- 0.5 < \cos i < 0.5$, that is, for $60^\circ < i < 120^\circ$.
This feature is the characteristic signature of the Lidov-Kozai secular perturbations, which periodically exchange inclination with eccentricity and thus facilitate the development of unstable orbits \citep[e.g.][]{1962P&SS....9..719L, 1962AJ.....67..591K}.
However, a small fraction of binary systems persist at the critical inclination range, corresponding to the shortest orbital periods, because general relativity apsidal precession breaks the Lidov-Kozai cycles  \citep{2007ApJ...669.1298F, 2016ARA&A..54..441N, 2018MNRAS.479.4749B}. 
This feature also explains why there is a peak of surviving systems for short orbital periods (Fig.~\ref{P_end}\,(a)).
For more details see Appendix~\ref{Appendix:Distortions} and \ref{Appendix:AnalyticalPredictions}.
A more subtle inspection of the inclination distribution shows that retrograde orbits ($\cos i < 0$) are also slightly more resilient than prograde orbits ($\cos i > 0$). 
This asymmetry can be explained owing to the distribution of mean motion resonances that overlap more easily and trigger chaos in the prograde configurations \citep[e.g.][]{2012MNRAS.424...52M}.

In Fig. \ref{P_end}\,(d), we show the distribution of the mass ratios.
Contrary to the previous cases, we observe that there is no mass preference for disruption apart from some statistical fluctuations.
Nevertheless, this figure gives us a clear view of the overall surviving fraction of binary systems ($\sim 30\%$).

In Fig.~\ref{Roche_limit_Fig}, we present the distribution of the surviving sample in the eccentricity versus orbital period plane, along with the corresponding distribution of surviving orbital periods. In the same plot, we also indicate the Roche limit. The observed increase in the number of surviving binaries between the orange and green regions is attributed to the greater availability of viable binary configurations permitted by the Roche limit.

\subsection{Impact of radial velocity constraints}
\label{K-section}

\begin{figure}
    \centering
    \includegraphics[width=0.95\columnwidth]{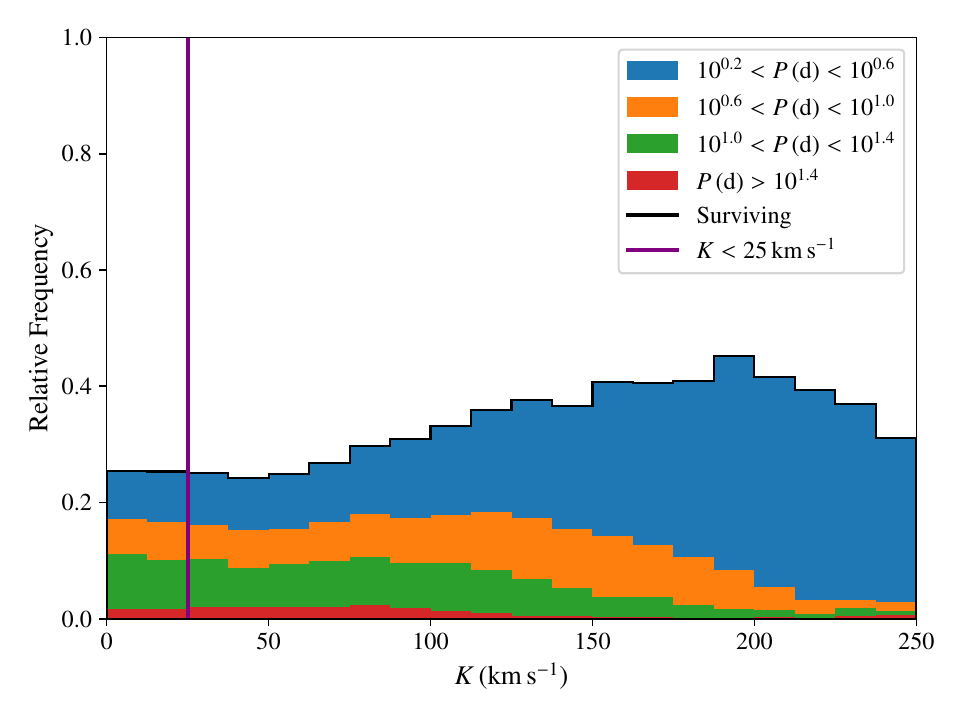}
    \caption{Histogram of the relative frequency distribution of the surviving binary's radial-velocity semi-amplitude, $K$ (Eq.\,(\ref{K-amplitude})). The solid line corresponds to the detection limit $K=25$~km/s \citep{2023ApJ...948...94C}. The color code corresponds to different ranges of orbital periods (Fig.~\ref{P_end}\,(a)).  \label{HistogramK}}
\end{figure}

\citet{2018ApJ...854...12C} ran a systematic search for spectroscopic binaries around the SMBH at the Galactic center.
This study includes radial velocity measurements of S2, showing no evidence of a binary companion with a semi-amplitude larger than 25~km/s. 
The radial velocity semi-amplitude of the binary, denoted as $K$, is given by \citep[e.g.][]{2010exop.book...15M}
\begin{equation}
K = \frac{m_{\mathrm{B}}}{m_{\mathrm{B}}+m_{\mathrm{A}}}\frac{n a \sin I}{\sqrt{1 - e^{2}}} = \frac{q  \sin I}{\left(1 + q\right)^{2/3}}\frac{(G m_\mathrm{A})^{1/3}}{\sqrt{1 - e^{2}}} \left(\frac{2 \pi}{P}\right)^{1/3}
\ , \label{K-amplitude}
\end{equation}
where $n = 2 \pi / P $ 
is the mean motion, $G$ is the gravitational constant, and $I$ is the inclination of the binary's orbital plane with respect to the plane of the sky.
This inclination can be obtained from our numerical simulations through the mutual inclination, $i$, and the longitude of the node, $\Omega$, as \citep[e.g.][]{2012A&A...541A.151G}
\begin{equation}
\cos I =  \cos i \cos I_{\bullet} - \sin i  \sin I_{\bullet} \cos \Omega \ , \label{cosI_formula}
\end{equation}
where $I_{\bullet}= 134.7^\circ $ is the inclination of S2 (Table~\ref{orb_elem_S2}).
We can thus determine which kind of systems would have remained undetected to the current date. 

In Fig.~\ref{P_end}, we additionally show the number of surviving stars with $K < 25 \, \rm km\,s^{-1}$  (purple line) as a function of the orbital parameters.
A first striking observation is that the fraction of undetectable systems is much lower than the number of surviving systems.
More precisely, only 4.4\% of the surviving binaries remain undetected through spectroscopic measurements.
This fraction therefore represents the probability that S2 is part of a binary system, given our initial set of conditions.
For completeness, in Fig.~\ref{HistogramK}, we show the distribution of the surviving systems as a function of the $K$ value.

\subsection{Properties of an eventual companion to S2}

A more detailed inspection of Fig.~\ref{P_end} also provides information on the type of binary system that can still be hidden, i.e., the fraction of surviving systems with $K<25$~km/s (purple line).

The orbital period distribution (Fig.~\ref{P_end}\,(a)) is nearly flat, there is only a slight bend for the period range $10^{0.6} < P < 10^{1.0}$~d (orange region).
Indeed, the excess of surviving systems with short orbital periods is balanced by the higher radial-velocity amplitudes of these systems, since $K \propto P^{-1/3}$  (Eq.\,(\ref{K-amplitude})).

The eccentricity distribution (Fig.~\ref{P_end}\,(b)) shows a slight increase for near-circular orbits, and the inclination distribution (Fig.~\ref{P_end}\,(c)) shows an excess for near-coplanar orbits. 
However, in both cases, the surplus is just a consequence of the higher number of surviving systems at those eccentricities and inclinations, respectively.
Therefore, we conclude that there is no selection effect for these two orbital parameters.

\begin{figure}
    \centering
    \includegraphics[width=0.95\columnwidth]{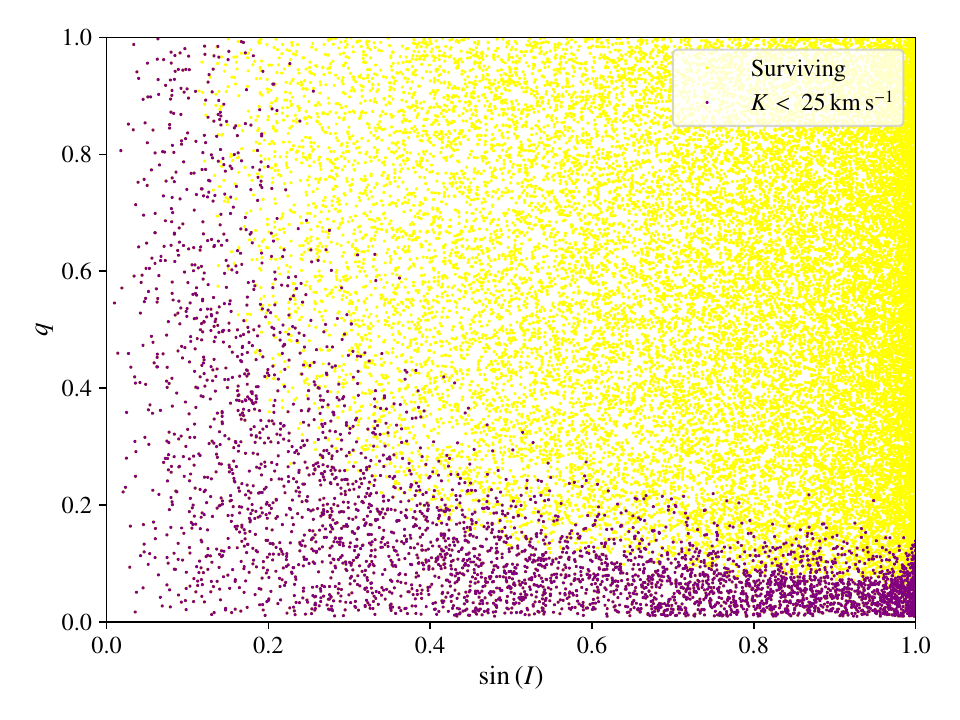}
    \caption{Distribution of the surviving binary systems as a function of the inclination to the line-of-sight, $I$, and the mass ratio, $q$. The surviving binaries with semi-amplitudes $K < 25 \, \rm km\,s^{-1}$ are given in purple, corresponding to the observational limitations. \label{i_sky_vs_q_end}}
\end{figure}

Concerning the companion mass (Fig.~\ref{P_end}\,(d)), we observe a clear preference for low-mass ratios, which is in agreement with Eq.\,(\ref{K-amplitude}), since $ K \propto q $.
In particular, we verify that the undetected binaries accumulate towards mass ratios $q < 0.15$, that is, for secondary masses $m \lesssim 2$~$M_\odot$.
Although much less likely, we cannot rule out secondary companions of any mass ratio.
Indeed, a closer look at those cases shows that they correspond to systems whose orbital plane is nearly aligned with the plane of the sky (see Fig.~\ref{i_sky_vs_q_end}), that is, for systems with $\sin I \ll 1$, also in accordance with Eq.\,(\ref{K-amplitude}).
However, the GRAVITY imaging data can detect any object near S2 that is brighter than approximately $18.5$~mag, corresponding to a main-sequence star with a mass of about $2 \, M_{\odot}$  \citep{2018A&A...615L..15G}. 
This aligns with \citet{2017ApJ...847..120H}, which found no double-line source.

\subsection{Astrometric signatures of surviving systems}

In a binary system, the motion of one star around the center of mass reveals wobbles caused by the gravitational pull of its companion. These shifts in position are also referred to as astrometric signals, whose semi-amplitude may be equated as (see Appendix~\ref{Appendix:AstrometricSignal})
\begin{equation}\label{astrometric_signal}
    \alpha = \frac{q}{\left(1 + q\right)^{2/3}}\frac{\left(G m_\mathrm{A} \right)^{1/3}}{n^{2/3} D} \left(\gamma  + \sqrt{\gamma^2 - (1-e^2) \cos^2 I} \, \right)^{1/2}\ ,
\end{equation}
where $D$ is the distance to the binary system,  
\begin{equation}
    \gamma = 1 - \frac{1}{2}\left[e^{2} 
 + \left(1-e^{2}\cos^2 \omega\right) \sin^2 I \, \right] \ ,
\end{equation}
and all other parameters retain their previous designations.

\begin{figure}
    \centering
    \includegraphics[width=0.95\columnwidth]{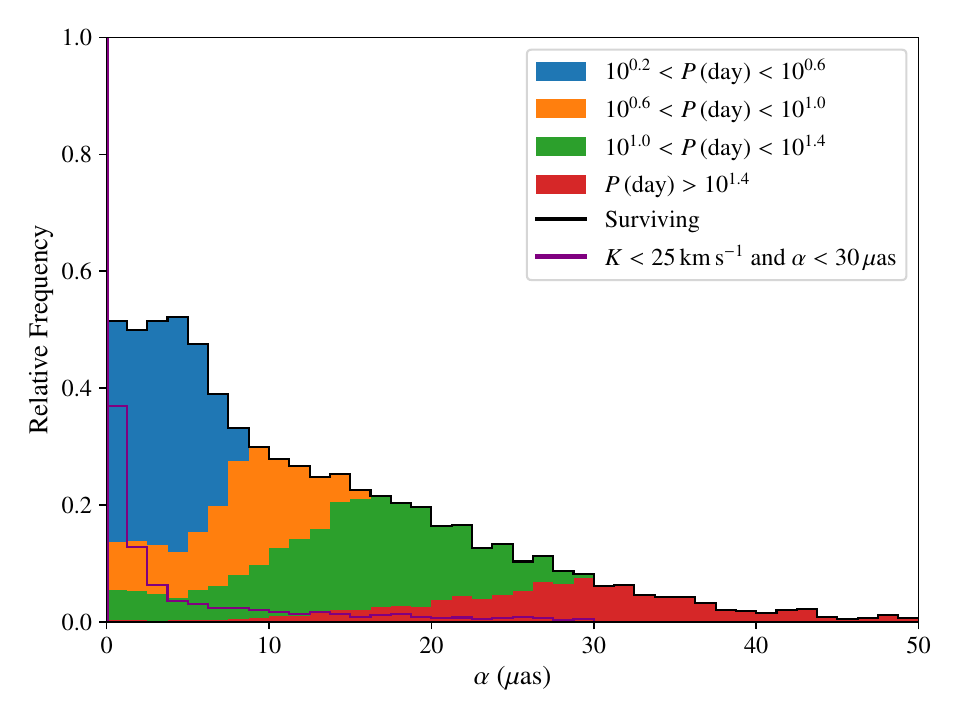}
    \caption{Histogram of the relative frequency distribution of the semi-amplitude of the astrometric signal for surviving binary systems, $\alpha$ (Eq.\,(\ref{astrometric_signal})). The solid purple line highlights cases where $K < 25 \, \rm km\,s^{-1}$ and $\alpha < 30 \, \mu\rm{as}$. The colour code corresponds to different ranges of orbital periods (Fig.~\ref{P_end}\,(a)). \label{Histogram_astrometric_signal}}
\end{figure}

In Fig.~\ref{Histogram_astrometric_signal}, we present the distribution of the astrometric wobble caused by a companion on S2. 
The current precision of the GRAVITY instrumentation is $\alpha \approx 30 \, \mu{\rm as}$ \citep{2018A&A...615L..15G}. 
Therefore, only a fraction of the long-period binaries can be detected (in red), the wobble caused by the vast majority of the binary companions is still out of reach.

In Fig.~\ref{Histogram_astrometric_signal}, we show the distribution of the semi-amplitude of the astrometric signal. Given the current observational uncertainties of GRAVITY, we find that a significant fraction of the surviving binary systems cannot yet be detected ($\alpha < 30 \, \mu{\rm as}$). In Fig.~\ref{Histogram_astrometric_signal}, the purple line outlines the subset of systems that remain undetected by both astrometric and radial velocity methods. We observe that these systems span a wide range of orbital periods, but are most prevalent for $P < 10$~d.

We estimate that the probability of a binary system remaining undetected by both astrometric and radial velocity measurements is approximately 4.3\%. This probability decreases further when we restrict ourselves to binaries that are undetectable through both methods and have companion masses lower than $ 2\,M_\odot$, yielding a probability of approximately 3.0\%.

\subsection{Constraints on the origin of S2}

The origin of the stars in the S-cluster remains elusive and different formation scenarios have been proposed.
One possibility is in situ formation of B-type and O-type stars in an accretion disk surrounding Sgr~A*, giving rise to a population of stellar-mass black holes. Subsequent dynamical interactions and collisions between these black holes and stars could account for the observed lack of O-type stars in the S-cluster, as well as the lack of hypervelocity stars (HVS) originating from the Galactic center \citep{2025A&A...695L..19H}.

Another scenario proposes the inward migration of stars originally formed in the nearby Nuclear Star Cluster (NSC) \cite[e.g.,][]{2007MNRAS.374..515L, 2010ApJ...709..597G, 2014ApJ...786L..14C}. Binary systems coming from outside the sphere of influence of Sgr~A* would then undergo tidal disruption, resulting in one star being captured by the SMBH and becoming a member of the S-cluster and another being ejected as an hypervelocity star (HVS) \cite[e.g.,][]{1988Natur.331..687H, 2006MNRAS.368..221G, 2008ApJ...683L.151L}.

At the end of our simulations, we obtain breakup and merger rates that are approximately 19.6\% and 50.6\%, respectively (Fig.~\ref{bin_frac}), that is, only 29.8\% of the initial binaries survive. While mergers only produce single stars, breakup events can lead to two possible outcomes: (1) one star is ejected as a HVS, while the other remains bounded to Sgr~A*; or (2) both stars remain bounded. The first scenario only accounts for approximately  0.01\% of the breakup events, yielding around 3~HVS in our simulations, so we conclude that breakup events most commonly originate two stars. In fact, several HVS have been detected that can be traced back to the GC \citep{2006ApJ...640L..35B}.

In the case of in situ formation, constraining the binary of stars in the S-cluster may provide insight into their origin.
About 70\% of the B-type stars in the Galactic field are found in binary systems \citep{2012Sci...337..444S}. 
Assuming the same initial fraction in case of in situ formation, after 1~Myr of evolution, our simulations predict that only 29.8\% of the 70\% remain as binary, and that 19.6\% of the 70\% produce two single stars. 
As a result, the final distribution expected would be about 81\% of the stars as single stars and the remaining 19\% as binary stars.
Although this corrected expected binary fraction is significantly lower than that observed in the Galactic field, it remains smaller than the currently measured binary fraction in the Galactic Center itself, which is approximately 42\% \citep{2023ApJ...948...94C}. Therefore, we conclude that in situ formation is unlikely to be the dominant scenario for the origin of these stars.

\subsection{Possibility of planetary companions}

The B-star Exoplanet Abundance Study (BEAST) was launched to investigate the frequency and properties of planets orbiting B-type stars in the Scorpius–Centaurus (Sco–Cen) association \citep{2019A&A...626A..99J}. This survey has already identified some systems composed by a B-type star and a planetary-mass companion \citep[e.g.,][]{2021Natur.600..231J, 2022A&A...664A...9S, 2023A&A...676L..10C, 2024A&A...692A.263D}.

\citet{2025AJ....169..131T} reported the detection of a hypervelocity star hosting a planet. Since these stars are believed to result from the Hills mechanism \citep[e.g.,][]{2003ApJ...599.1129Y}, it is plausible to hypothesize that the inverse scenario also occurs, that is, the star–planet system remains bounded and orbiting Sgr~A*.

In Fig.~\ref{P_end}\,(d), we observe that the mass ratio of the surviving binaries is uniformly distributed and does not exhibit any preferential range. 
Thus, companion masses with $q < 0.01$ should not modify the dynamics of the problem.
As a result, we expect that the distributions of the orbital period, eccentricity, and mutual inclination (Fig.~\ref{P_end}) remain essentially unchanged and valid also for planetary companions around the S2-star.
These planetary companions would not be possible to observe with the current radial velocity or astrometric measurements, explaining why such detections did not occur yet. However, the detection could be made possible using the transits method \citep{2012MNRAS.423..948G}.

Planets around stars in the S-cluster may also be engulfed (merger) or ejected (breakup). 
In the latter case, we end up with two possible new interesting scenarios:  (1) the emission of hypervelocity planets; or (2) the planets settle in orbits around Sgr~A*, resulting in a new population of objects at the Galactic center. Due to the extremely faint nature of the planets, the confirmation of both scenarios is still beyond the capabilities of current observational techniques.

\subsection{Limitations of the present work}

The primary limitations of this study include the simulation timescale, the three-body system approximation, and the neglect of potential interactions of S2 with some undetected stars. A simulation time of $10^{6}$~yr excludes dynamical processes that act on longer timescales, such as scalar and vector resonant relaxation \citep[see Fig.~1 in][]{2011MNRAS.412..187K}. These relaxation processes can induce torques on binaries, thereby affecting their survival probability \citep{2015MNRAS.448.3265K}.
The three-body approximation does not account for perturbations from other members of the S-cluster or for the influence of the clockwise stellar disk. Neglecting potential collisions between S2 and other stars, the model may underestimate the rate of binary disruption events. Finally, we adopt a conservative merger criterion, namely the Roche limit, rather than Eggleton's formula \citep{1983ApJ...268..368E}. 
A fraction of merged systems might survive as binaries when employing Eggleton's formula, but that would require modeling eccentric mass transfer (and stellar evolution), which is inherently challenging and lies beyond the scope of the present study.

\section{Summary and conclusions}\label{conclusions}

In this paper, we aim to constrain the possible presence of companions to the S2 star, which are strongly perturbed by the presence of the SMBH Sgr~A*. 
Using a Monte Carlo approach, we adopted uniform distributions of the initial orbital parameters, and through $N$-body simulations we explored which orbital configurations are dynamically stable and can survive in such a hostile environment.

We find that stable binary systems are restricted to orbital periods shorter than 100 days and eccentricities below 0.8. Furthermore, while companions can exist across all mutual inclinations, they are more likely to survive in near-coplanar configurations. In particular, for mutual inclinations between $60^\circ$ and $120^\circ$, only binaries with orbital periods shorter than 4 days remain gravitationally bounded. This behavior arises because Lidov–Kozai cycles are suppressed by general relativity apsidal precession in such close-in orbits.

Additionally, we find that the mass ratio distribution is uniformly disrupted; that is, this parameter is not critical for the stability. As a result, we can extrapolate our results to systems with mass ratios consistent with planetary companions. We thus conclude that the S2 star can harbor a planetary companion with an orbital period shorter than 100 days and predominantly in a coplanar orbit.

Furthermore, our simulations also allow us to track mergers and tidal breakup events. Mergers are most commonly observed in systems with mutual inclinations between $60^\circ$ and $120^\circ$, triggered by Lidov-Kozai oscillations. Tidal breakups predominantly occur in binaries with initial orbital periods longer than 20 days, as these systems are more susceptible to disruption near the Hill sphere of S2. Among the outcome of tidal breakup events, we have counted three hypervelocity objects.

Observational studies such as \citet{2018ApJ...854...12C, 2023ApJ...948...94C} have used radial velocity measurements to place constraints on the possible mass of a companion to S2. However, due to observational limitations, these studies do not provide information about the orbital configuration of an undetected binary. In this work, we find that such configurations are restricted to orbital periods shorter than 100~days, eccentricities primarily below 0.6, and nearly coplanar orbits. The undetected systems are also expected to have low mass ratios ($q \le 0.15$).

A possible companion to S2 would induce reflex motion around the system's center of mass, thereby biasing the S2's astrometric signal. The commonly used expression for computing the semi-amplitude of the astrometric signal assumes a circular, face-on orbit with respect to the plane of the sky, an idealized scenario. In reality, most orbits exhibit non-zero inclination and eccentricity. In this paper, we derive the projected semi-amplitude of the astrometric signal for orbits with arbitrary eccentricity and inclination. We find that the large majority of the surviving systems have an astrometric signature $\lesssim 30 \, \mu\mathrm{as}$, which is still below the current detection limits of the GRAVITY instrumentation \citep{2018A&A...615L..15G}.

Finally, given the current radial velocity detection threshold of approximately 25~km\,s$^{-1}$ \citep{2023ApJ...948...94C}, we estimate a 4.4\% probability that S2 remains part of an undetected binary system. When incorporating both radial velocity and astrometric constraints, this probability reduces to 4.3\%. Moreover, if we further restrict our analysis to undetected binaries with companion masses lower than $ 2\,M_\odot$, the probability of S2 being part of a binary system is reduced to 3.0\%.

\begin{acknowledgements}
We thank the anonymous reviewer for several suggestions that improved the manuscript. We also thank Nuno Morujão and Tiago Gomes for their support in using computational resources, and Stefan Gillessen for valuable discussions.
This work was financed through national funds by FCT - Funda\c{c}\~ao para a Ci\^encia e a Tecnologia, I.P., Portugal, in the framework of the projects
2024.01252.BD, CFisUC UID/04564/2025 (with DOI identifier 10.54499/UID/04564/2025), and to the Center for Astrophysics and Gravitation (CENTRA/IST/ULisboa) through grant No. UID/99/2025.
TB is supported by the European Union’s Horizon Europe research and innovation
programme under the Marie Skłodowska–Curie grant agreement No 101153423.
We acknowledge the Laboratory for Advanced Computing at the University of Coimbra (\href{https://www.uc.pt/lca}{https://www.uc.pt/lca}) and Rede Nacional de Computação Avançada, under grants 2025.00007.HPCVLAB.UPORTO and 2025.08956.CPCA.A1, for providing the resources to perform the numerical simulations.
\end{acknowledgements}

\bibliographystyle{aa} 
\bibliography{aa56945-25} 

@ARTICLE{1983ApJ...268..368E,
       author = {{Eggleton}, P.~P.},
        title = "{Aproximations to the radii of Roche lobes.}",
      journal = {\apj},
         year = 1983,
        month = may,
       volume = {268},
        pages = {368-369},
          doi = {10.1086/160960},
       adsurl = {https://ui.adsabs.harvard.edu/abs/1983ApJ...268..368E}
}

@ARTICLE{2009MNRAS.394.1085T,
       author = {{Touma}, J.~R. and {Tremaine}, S. and {Kazandjian}, M.~V.},
        title = "{Gauss's method for secular dynamics, softened}",
      journal = {\mnras},
         year = 2009,
        month = apr,
       volume = {394},
       number = {2},
        pages = {1085-1108},
          doi = {10.1111/j.1365-2966.2009.14409.x},
archivePrefix = {arXiv},
       eprint = {0811.2812},
 primaryClass = {astro-ph},
       adsurl = {https://ui.adsabs.harvard.edu/abs/2009MNRAS.394.1085T}
}

@ARTICLE{2024ApJ...962...81B,
       author = {{Burkert}, A. and {Gillessen}, S. and {Lin}, D.~N.~C. and {Zheng}, X. and {Schoeller}, P. and {Eisenhauer}, F. and {Genzel}, R.},
        title = "{The Orbital Structure and Selection Effects of the Galactic Center S-star Cluster}",
      journal = {\apj},
         year = 2024,
        month = feb,
       volume = {962},
       number = {1},
          eid = {81},
        pages = {81},
          doi = {10.3847/1538-4357/ad17bb},
archivePrefix = {arXiv},
       eprint = {2306.02076},
 primaryClass = {astro-ph.GA},
       adsurl = {https://ui.adsabs.harvard.edu/abs/2024ApJ...962...81B}
}

@ARTICLE{2015MNRAS.448.3265K,
       author = {{Kocsis}, Bence and {Tremaine}, Scott},
        title = "{A numerical study of vector resonant relaxation}",
      journal = {\mnras},
         year = 2015,
        month = apr,
       volume = {448},
       number = {4},
        pages = {3265-3296},
          doi = {10.1093/mnras/stv057},
archivePrefix = {arXiv},
       eprint = {1406.1178},
 primaryClass = {astro-ph.GA},
       adsurl = {https://ui.adsabs.harvard.edu/abs/2015MNRAS.448.3265K}
}

@ARTICLE{2023ApJ...955...30R,
       author = {{Rose}, Sanaea C. and {Naoz}, Smadar and {Sari}, Re'em and {Linial}, Itai},
        title = "{Stellar Collisions in the Galactic Center: Massive Stars, Collision Remnants, and Missing Red Giants}",
      journal = {\apj},
         year = 2023,
        month = sep,
       volume = {955},
       number = {1},
          eid = {30},
        pages = {30},
          doi = {10.3847/1538-4357/acee75},
archivePrefix = {arXiv},
       eprint = {2304.10569},
 primaryClass = {astro-ph.GA},
       adsurl = {https://ui.adsabs.harvard.edu/abs/2023ApJ...955...30R}
}

@ARTICLE{2002Natur.419..694S,
       author = {{Sch{\"o}del}, R. and {Ott}, T. and {Genzel}, R. and {Hofmann}, R. and {Lehnert}, M. and {Eckart}, A. and {Mouawad}, N. and {Alexander}, T. and {Reid}, M.~J. and {Lenzen}, R. and {Hartung}, M. and {Lacombe}, F. and {Rouan}, D. and {Gendron}, E. and {Rousset}, G. and {Lagrange}, A. -M. and {Brandner}, W. and {Ageorges}, N. and {Lidman}, C. and {Moorwood}, A.~F.~M. and {Spyromilio}, J. and {Hubin}, N. and {Menten}, K.~M.},
        title = "{A star in a 15.2-year orbit around the supermassive black hole at the centre of the Milky Way}",
      journal = {\nat},
         year = 2002,
        month = oct,
       volume = {419},
       number = {6908},
        pages = {694-696},
          doi = {10.1038/nature01121},
archivePrefix = {arXiv},
       eprint = {astro-ph/0210426},
 primaryClass = {astro-ph},
       adsurl = {https://ui.adsabs.harvard.edu/abs/2002Natur.419..694S}
}

@ARTICLE{2013ARA&A..51..269D,
       author = {{Duch{\^e}ne}, Gaspard and {Kraus}, Adam},
        title = "{Stellar Multiplicity}",
      journal = {\araa},
         year = 2013,
        month = aug,
       volume = {51},
       number = {1},
        pages = {269-310},
          doi = {10.1146/annurev-astro-081710-102602},
archivePrefix = {arXiv},
       eprint = {1303.3028},
 primaryClass = {astro-ph.SR},
       adsurl = {https://ui.adsabs.harvard.edu/abs/2013ARA&A..51..269D}
}

@ARTICLE{2010RvMP...82.3121G,
       author = {{Genzel}, Reinhard and {Eisenhauer}, Frank and {Gillessen}, Stefan},
        title = "{The Galactic Center massive black hole and nuclear star cluster}",
      journal = {Reviews of Modern Physics},
         year = 2010,
        month = oct,
       volume = {82},
       number = {4},
        pages = {3121-3195},
          doi = {10.1103/RevModPhys.82.3121},
archivePrefix = {arXiv},
       eprint = {1006.0064},
 primaryClass = {astro-ph.GA},
       adsurl = {https://ui.adsabs.harvard.edu/abs/2010RvMP...82.3121G}
}

@ARTICLE{2018ApJ...853L..24N,
       author = {{Naoz}, Smadar and {Ghez}, Andrea M. and {Hees}, Aurelien and {Do}, Tuan and {Witzel}, Gunther and {Lu}, Jessica R.},
        title = "{Confusing Binaries: The Role of Stellar Binaries in Biasing Disk Properties in the Galactic Center}",
      journal = {\apjl},
         year = 2018,
        month = feb,
       volume = {853},
       number = {2},
          eid = {L24},
        pages = {L24},
          doi = {10.3847/2041-8213/aaa6bf},
archivePrefix = {arXiv},
       eprint = {1801.03934},
 primaryClass = {astro-ph.GA},
       adsurl = {https://ui.adsabs.harvard.edu/abs/2018ApJ...853L..24N}
}

@ARTICLE{2024ApJ...964..164G,
       author = {{Gautam}, Abhimat K. and {Do}, Tuan and {Ghez}, Andrea M. and {Chu}, Devin S. and {Hosek}, Matthew W. and {Sakai}, Shoko and {Naoz}, Smadar and {Morris}, Mark R. and {Ciurlo}, Anna and {Haggard}, Zo{\"e} and {Lu}, Jessica R.},
        title = "{An Estimate of the Binary Star Fraction among Young Stars at the Galactic Center: Possible Evidence of a Radial Dependence}",
      journal = {\apj},
         year = 2024,
        month = apr,
       volume = {964},
       number = {2},
          eid = {164},
        pages = {164},
          doi = {10.3847/1538-4357/ad26e6},
archivePrefix = {arXiv},
       eprint = {2401.12555},
 primaryClass = {astro-ph.GA},
       adsurl = {https://ui.adsabs.harvard.edu/abs/2024ApJ...964..164G}
}

@ARTICLE{1997MNRAS.284..576E,
       author = {{Eckart}, A. and {Genzel}, R.},
        title = "{Stellar proper motions in the central 0.1 PC of the Galaxy}",
      journal = {\mnras},
         year = 1997,
        month = jan,
       volume = {284},
       number = {3},
        pages = {576-598},
          doi = {10.1093/mnras/284.3.576},
       adsurl = {https://ui.adsabs.harvard.edu/abs/1997MNRAS.284..576E}
}

@ARTICLE{2017A&A...602A..94G,
       author = {{GRAVITY Collaboration} and {Abuter}, R. and {Accardo}, M. and {Amorim}, A. and {Anugu}, N. and {{\'A}vila}, G. and {Azouaoui}, N. and {Benisty}, M. and {Berger}, J.~P. and {Blind}, N. and {Bonnet}, H. and {Bourget}, P. and {Brandner}, W. and {Brast}, R. and {Buron}, A. and {Burtscher}, L. and {Cassaing}, F. and {Chapron}, F. and {Choquet}, {\'E}. and {Cl{\'e}net}, Y. and {Collin}, C. and {Coud{\'e} Du Foresto}, V. and {de Wit}, W. and {de Zeeuw}, P.~T. and {Deen}, C. and {Delplancke-Str{\"o}bele}, F. and {Dembet}, R. and {Derie}, F. and {Dexter}, J. and {Duvert}, G. and {Ebert}, M. and {Eckart}, A. and {Eisenhauer}, F. and {Esselborn}, M. and {F{\'e}dou}, P. and {Finger}, G. and {Garcia}, P. and {Garcia Dabo}, C.~E. and {Garcia Lopez}, R. and {Gendron}, E. and {Genzel}, R. and {Gillessen}, S. and {Gonte}, F. and {Gordo}, P. and {Grould}, M. and {Gr{\"o}zinger}, U. and {Guieu}, S. and {Haguenauer}, P. and {Hans}, O. and {Haubois}, X. and {Haug}, M. and {Haussmann}, F. and {Henning}, Th. and {Hippler}, S. and {Horrobin}, M. and {Huber}, A. and {Hubert}, Z. and {Hubin}, N. and {Hummel}, C.~A. and {Jakob}, G. and {Janssen}, A. and {Jochum}, L. and {Jocou}, L. and {Kaufer}, A. and {Kellner}, S. and {Kendrew}, S. and {Kern}, L. and {Kervella}, P. and {Kiekebusch}, M. and {Klein}, R. and {Kok}, Y. and {Kolb}, J. and {Kulas}, M. and {Lacour}, S. and {Lapeyr{\`e}re}, V. and {Lazareff}, B. and {Le Bouquin}, J. -B. and {L{\`e}na}, P. and {Lenzen}, R. and {L{\'e}v{\^e}que}, S. and {Lippa}, M. and {Magnard}, Y. and {Mehrgan}, L. and {Mellein}, M. and {M{\'e}rand}, A. and {Moreno-Ventas}, J. and {Moulin}, T. and {M{\"u}ller}, E. and {M{\"u}ller}, F. and {Neumann}, U. and {Oberti}, S. and {Ott}, T. and {Pallanca}, L. and {Panduro}, J. and {Pasquini}, L. and {Paumard}, T. and {Percheron}, I. and {Perraut}, K. and {Perrin}, G. and {Pfl{\"u}ger}, A. and {Pfuhl}, O. and {Phan Duc}, T. and {Plewa}, P.~M. and {Popovic}, D. and {Rabien}, S. and {Ram{\'\i}rez}, A. and {Ramos}, J. and {Rau}, C. and {Riquelme}, M. and {Rohloff}, R. -R. and {Rousset}, G. and {Sanchez-Bermudez}, J. and {Scheithauer}, S. and {Sch{\"o}ller}, M. and {Schuhler}, N. and {Spyromilio}, J. and {Straubmeier}, C. and {Sturm}, E. and {Suarez}, M. and {Tristram}, K.~R.~W. and {Ventura}, N. and {Vincent}, F. and {Waisberg}, I. and {Wank}, I. and {Weber}, J. and {Wieprecht}, E. and {Wiest}, M. and {Wiezorrek}, E. and {Wittkowski}, M. and {Woillez}, J. and {Wolff}, B. and {Yazici}, S. and {Ziegler}, D. and {Zins}, G.},
        title = "{First light for GRAVITY: Phase referencing optical interferometry for the Very Large Telescope Interferometer}",
      journal = {\aap},
         year = 2017,
        month = jun,
       volume = {602},
          eid = {A94},
        pages = {A94},
          doi = {10.1051/0004-6361/201730838},
archivePrefix = {arXiv},
       eprint = {1705.02345},
 primaryClass = {astro-ph.IM},
       adsurl = {https://ui.adsabs.harvard.edu/abs/2017A&A...602A..94G}
}

@ARTICLE{2003ApJ...586L.127G,
       author = {{Ghez}, A.~M. and {Duch{\^e}ne}, G. and {Matthews}, K. and {Hornstein}, S.~D. and {Tanner}, A. and {Larkin}, J. and {Morris}, M. and {Becklin}, E.~E. and {Salim}, S. and {Kremenek}, T. and {Thompson}, D. and {Soifer}, B.~T. and {Neugebauer}, G. and {McLean}, I.},
        title = "{The First Measurement of Spectral Lines in a Short-Period Star Bound to the Galaxy's Central Black Hole: A Paradox of Youth}",
      journal = {\apjl},
         year = 2003,
        month = apr,
       volume = {586},
       number = {2},
        pages = {L127-L131},
          doi = {10.1086/374804},
archivePrefix = {arXiv},
       eprint = {astro-ph/0302299},
 primaryClass = {astro-ph},
       adsurl = {https://ui.adsabs.harvard.edu/abs/2003ApJ...586L.127G}
}

@ARTICLE{2008ApJ...672L.119M,
       author = {{Martins}, F. and {Gillessen}, S. and {Eisenhauer}, F. and {Genzel}, R. and {Ott}, T. and {Trippe}, S.},
        title = "{On the Nature of the Fast-Moving Star S2 in the Galactic Center}",
      journal = {\apjl},
         year = 2008,
        month = jan,
       volume = {672},
       number = {2},
        pages = {L119},
          doi = {10.1086/526768},
archivePrefix = {arXiv},
       eprint = {0711.3344},
 primaryClass = {astro-ph},
       adsurl = {https://ui.adsabs.harvard.edu/abs/2008ApJ...672L.119M}
}

@ARTICLE{2017arXiv171106389D,
       author = {{Do}, Tuan and {Hees}, Aurelien and {Dehghanfar}, Arezu and {Ghez}, Andrea and {Wright}, Shelley},
        title = "{Measuring the effects of General Relativity at the Galactic Center with Future Extremely Large Telescopes}",
      journal = {arXiv e-prints},
         year = 2017,
        month = nov,
          eid = {arXiv:1711.06389},
        pages = {arXiv:1711.06389},
          doi = {10.48550/arXiv.1711.06389},
archivePrefix = {arXiv},
       eprint = {1711.06389},
 primaryClass = {astro-ph.GA},
       adsurl = {https://ui.adsabs.harvard.edu/abs/2017arXiv171106389D}
}

@ARTICLE{2022Msngr.189...17A,
       author = {{GRAVITY+ Collaboration} and {Abuter}, R. and {Alarcon}, P. and {Allouche}, F. and {Amorim}, A. and {Bailet}, C. and {Bedigan}, H. and {Berdeu}, A. and {Berger}, J. -P. and {Berio}, P. and {Bigioli}, A. and {Blaho}, R. and {Boebion}, O. and {Bolzer}, M. -L. and {Bonnet}, H. and {Bourdarot}, G. and {Bourget}, P. and {Brandner}, W. and {Cardenas}, C. and {Conzelmann}, R. and {Comin}, M. and {Cl{\'e}net}, Y. and {Courtney-Barrer}, B. and {Dallilar}, Y. and {Davies}, R. and {Defr{\`e}re}, D. and {Delboulb{\'e}}, A. and {Delplancke-Str{\"o}bele}, F. and {Dembet}, R. and {de Zeeuw}, T. and {Drescher}, A. and {Eckart}, A. and {{\'E}douard}, C. and {Eisenhauer}, F. and {Fabricius}, M. and {Feuchtgruber}, H. and {Finger}, G. and {F{\"o}rster Schreiber}, N.~M. and {Fuenteseca}, E. and {Garcia}, E. and {Garcia}, P. and {Gao}, F. and {Gendron}, E. and {Genzel}, R. and {Gil}, J.~P. and {Gillessen}, S. and {Gomes}, T. and {Gont{\'e}}, F. and {Gouvret}, C. and {Guajardo}, P. and {Guidolin}, I. and {Guieu}, S. and {Guzmann}, R. and {Hackenberg}, W. and {Haddad}, N. and {Hartl}, M. and {Haubois}, X. and {Hau{\ss}mann}, F. and {Hei{\ss}el}, G. and {Henning}, T. and {Hippler}, S. and {H{\"o}nig}, S. and {Horrobin}, M. and {Hubin}, N. and {Jacqmart}, E. and {Jocou}, L. and {Kaufer}, A. and {Kervella}, P. and {Kirchbauer}, J. -P. and {Kolb}, J. and {Korhonen}, H. and {Kreidberg}, L. and {Krempl}, P. and {Lacour}, S. and {Lagarde}, S. and {Lai}, O. and {Lapeyr{\`e}re}, V. and {Laugier}, R. and {Le Bouquin}, J. -B. and {Leftley}, J. and {L{\'e}na}, P. and {Lewis}, S. and {Lutz}, D. and {Magnard}, Y. and {Mang}, F. and {Marcotto}, A. and {Maurel}, D. and {M{\'e}rand}, A. and {Millour}, F. and {More}, N. and {Nowacki}, H. and {Nowak}, M. and {Oberti}, S. and {Olivares}, F. and {Ott}, T. and {Pallanca}, L. and {Paumard}, T. and {Perraut}, K. and {Perrin}, G. and {Petrov}, R. and {Pfuhl}, O. and {Pourr{\'e}}, N. and {Rabien}, S. and {Rau}, C. and {Riquelme}, M. and {Robbe-Dubois}, S. and {Rochat}, S. and {Salman}, M. and {Scherbarth}, M. and {Sch{\"o}ller}, M. and {Schubert}, J. and {Schuhler}, N. and {Shangguan}, J. and {Shchekaturov}, P. and {Shimizu}, T. and {Scheithauer}, S. and {Sevin}, A. and {Soenke}, C. and {Soulez}, F. and {Spang}, A. and {Stadler}, E. and {Straubmeier}, C. and {Sturm}, E. and {Sykes}, C. and {Tacconi}, L. and {Tischer}, H. and {Tristram}, K. and {Vincent}, F. and {von Fellenberg}, S. and {Uysal}, S. and {Widmann}, F. and {Wieprecht}, E. and {Wiezorrek}, E. and {Woillez}, J. and {Yaz{\i}c{\i}}, {\c{S}}. and {Zins}, G.},
        title = "{The GRAVITY+ Project: Towards All-sky, Faint-Science, High-Contrast Near-Infrared Interferometry at the VLTI}",
      journal = {The Messenger},
         year = 2022,
        month = dec,
       volume = {189},
        pages = {17-22},
          doi = {10.18727/0722-6691/5285},
archivePrefix = {arXiv},
       eprint = {2301.08071},
 primaryClass = {astro-ph.IM},
       adsurl = {https://ui.adsabs.harvard.edu/abs/2022Msngr.189...17A}
}

@INPROCEEDINGS{2024SPIE13096E..11S,
       author = {{Sturm}, E. and {Davies}, R. and {Alves}, J. and {Cl{\'e}net}, Y. and {Kotilainen}, J. and {Monna}, A. and {Nicklas}, H. and {Pott}, J. -U. and {Tolstoy}, E. and {Vulcani}, B. and {Achren}, J. and {Annadevara}, S. and {Anwand-Heerwart}, H. and {Arcidiacono}, C. and {Barboza}, S. and {Barl}, L. and {Baudoz}, P. and {Bender}, R. and {Bezawada}, N. and {Biondi}, F. and {Bizenberger}, P. and {Blin}, A. and {Bon{\'e}}, A. and {Bonifacio}, P. and {Borgo}, B. and {Born}, J. van den and {Buey}, T. and {Cao}, Y. and {Chapron}, F. and {Chauvin}, G. and {Chemla}, F. and {Cloiseau}, K. and {Cohen}, M. and {Colin}, C. and {Czoske}, O. and {Dette}, J. -O. and {Deysenroth}, M. and {Dijkstra}, E. and {Dreizler}, S. and {Dupuis}, O. and {Egmond}, G. van and {Eisenhauer}, F. and {Elswijk}, E. and {Emslander}, A. and {Fabricius}, M. and {Fasola}, G. and {Ferreira}, F. and {F{\"o}rster Schreiber}, N. and {Fontana}, A. and {Gaudemard}, J. and {Gautherot}, N. and {Gendron}, E. and {Gennet}, C. and {Genzel}, R. and {Ghouchou}, L. and {Gillessen}, S. and {Gratadour}, D. and {Grazian}, A. and {Grupp}, F. and {Guieu}, S. and {Gullieuszik}, M. and {Haan}, M. de and {Hartke}, J. and {Hartl}, M. and {Haussmann}, F. and {Helin}, T. and {Hess}, H. -J. and {Hofferbert}, R. and {Huber}, H. and {Huby}, E. and {Huet}, J. -M. and {Ives}, D. and {Janssen}, A. and {Jaufmann}, P. and {Jilg}, T. and {Jodlbauer}, D. and {Jost}, J. and {Kausch}, W. and {Kellermann}, H. and {Kerber}, F. and {Kravcar}, H. and {Kravchenko}, K. and {Kulcs{\'a}r}, C. and {Kuncarayakti}, H. and {Kunst}, P. and {Kwast}, S. and {Lang}, F. and {Lange}, J. and {Lapeyrere}, V. and {Le Ruyet}, B. and {Leschinski}, K. and {Locatelli}, H. and {Massari}, D. and {Mattila}, S. and {Mei}, S. and {Merlin}, F. and {Meyer}, E. and {Michel}, C. and {Mohr}, L. and {Montarg{\`e}s}, M. and {M{\"u}ller}, F. and {M{\"u}nch}, N. and {Navarro}, R. and {Neumann}, U. and {Neumayer}, N. and {Neumeier}, L. and {Pedichini}, F. and {Pfl{\"u}ger}, A. and {Piazzesi}, R. and {Pinard}, L. and {Porras}, J. and {Portulari}, E. and {Przybilla}, N. and {Rabien}, S. and {Raffard}, J. and {Ragazzoni}, R. and {Ramlau}, R. and {Ramos}, J. and {Ramsay}, S. and {Raynaud}, H. -F. and {Rhode}, P. and {Richter}, A. and {Rix}, H. -W. and {Rodenhuis}, M. and {Rohloff}, R. -R. and {Romp}, R. and {Rousselot}, P. and {Sabha}, N. and {Sassolas}, B. and {Schlichter}, J. and {Schuil}, M. and {Schweitzer}, M. and {Seemann}, U. and {Sevin}, A. and {Simioni}, M. and {Spallek}, L. and {S{\"o}nmez}, A. and {Suuronen}, J. and {Taburet}, S. and {Thomas}, J. and {Tisserand}, E. and {Vaccari}, P. and {Valenti}, E. and {Verdoes Kleijn}, G. and {Verdugo}, M. and {Vidal}, F. and {Wagner}, R. and {Wegner}, M. and {Winden}, D. van and {Witschel}, J. and {Zanella}, A. and {Zeilinger}, W. and {Ziegleder}, J. and {Ziegler}, B.},
        title = "{The MICADO first light imager for the ELT: overview and current status}",
    booktitle = {Ground-based and Airborne Instrumentation for Astronomy X},
         year = 2024,
       editor = {{Bryant}, Julia J. and {Motohara}, Kentaro and {Vernet}, Jo{\"e}l. R.~D.},
       series = {Society of Photo-Optical Instrumentation Engineers (SPIE) Conference Series},
       volume = {13096},
        month = jul,
          eid = {1309611},
        pages = {1309611},
          doi = {10.1117/12.3017752},
       adsurl = {https://ui.adsabs.harvard.edu/abs/2024SPIE13096E..11S}
}

@ARTICLE{2023RPPh...86j4901D,
       author = {{De Laurentis}, Mariafelicia and {de Martino}, Ivan and {Della Monica}, Riccardo},
        title = "{The Galactic Center as a laboratory for theories of gravity and dark matter}",
      journal = {Reports on Progress in Physics},
         year = 2023,
        month = oct,
       volume = {86},
       number = {10},
          eid = {104901},
        pages = {104901},
          doi = {10.1088/1361-6633/ace91b},
archivePrefix = {arXiv},
       eprint = {2211.07008},
 primaryClass = {astro-ph.GA},
       adsurl = {https://ui.adsabs.harvard.edu/abs/2023RPPh...86j4901D}
}

@ARTICLE{2024MNRAS.530.3740G,
       author = {{GRAVITY Collaboration} and {Foschi}, A. and {Abuter}, R. and {Abd El Dayem}, K. and {Aimar}, N. and {Amaro Seoane}, P. and {Amorim}, A. and {Berger}, J.~P. and {Bonnet}, H. and {Bourdarot}, G. and {Brandner}, W. and {Davies}, R. and {de Zeeuw}, P.~T. and {Defr{\`e}re}, D. and {Dexter}, J. and {Drescher}, A. and {Eckart}, A. and {Eisenhauer}, F. and {F{\"o}rster Schreiber}, N.~M. and {Garcia}, P.~J.~V. and {Genzel}, R. and {Gillessen}, S. and {Gomes}, T. and {Haubois}, X. and {Hei{\ss}el}, G. and {Henning}, Th and {Jochum}, L. and {Jocou}, L. and {Kaufer}, A. and {Kreidberg}, L. and {Lacour}, S. and {Lapeyr{\`e}re}, V. and {Le Bouquin}, J. -B. and {L{\'e}na}, P. and {Lutz}, D. and {Mang}, F. and {Millour}, F. and {Ott}, T. and {Paumard}, T. and {Perraut}, K. and {Perrin}, G. and {Pfuhl}, O. and {Rabien}, S. and {Ribeiro}, D.~C. and {Sadun Bordoni}, M. and {Scheithauer}, S. and {Shangguan}, J. and {Shimizu}, T. and {Stadler}, J. and {Straubmeier}, C. and {Sturm}, E. and {Subroweit}, M. and {Tacconi}, L.~J. and {Vincent}, F. and {von Fellenberg}, S. and {Woillez}, J.},
        title = "{Using the motion of S2 to constrain vector clouds around Sgr A*}",
      journal = {\mnras},
         year = 2024,
        month = jun,
       volume = {530},
       number = {4},
        pages = {3740-3751},
          doi = {10.1093/mnras/stae423},
       adsurl = {https://ui.adsabs.harvard.edu/abs/2024MNRAS.530.3740G}
}

@ARTICLE{2023PhRvD.108j1303D,
       author = {{Della Monica}, Riccardo and {de Martino}, Ivan},
        title = "{Bounding the mass of ultralight bosonic dark matter particles with the motion of the S2 star around Sgr A$^{*}$}",
      journal = {\prd},
         year = 2023,
        month = nov,
       volume = {108},
       number = {10},
          eid = {L101303},
        pages = {L101303},
          doi = {10.1103/PhysRevD.108.L101303},
archivePrefix = {arXiv},
       eprint = {2305.10242},
 primaryClass = {gr-qc},
       adsurl = {https://ui.adsabs.harvard.edu/abs/2023PhRvD.108j1303D}
}

@ARTICLE{2024MNRAS.527.3196S,
       author = {{Shen}, Zhao-Qiang and {Yuan}, Guan-Wen and {Jiang}, Cheng-Zi and {Tsai}, Yue-Lin Sming and {Yuan}, Qiang and {Fan}, Yi-Zhong},
        title = "{Exploring dark matter spike distribution around the Galactic centre with stellar orbits}",
      journal = {\mnras},
         year = 2024,
        month = jan,
       volume = {527},
       number = {2},
        pages = {3196-3207},
          doi = {10.1093/mnras/stad3282},
archivePrefix = {arXiv},
       eprint = {2303.09284},
 primaryClass = {astro-ph.GA},
       adsurl = {https://ui.adsabs.harvard.edu/abs/2024MNRAS.527.3196S}
}

@ARTICLE{2021A&A...647A..59G,
       author = {{GRAVITY Collaboration} and {Abuter}, R. and {Amorim}, A. and {Baub{\"o}ck}, M. and {Berger}, J.~P. and {Bonnet}, H. and {Brandner}, W. and {Cl{\'e}net}, Y. and {Davies}, R. and {de Zeeuw}, P.~T. and {Dexter}, J. and {Dallilar}, Y. and {Drescher}, A. and {Eckart}, A. and {Eisenhauer}, F. and {F{\"o}rster Schreiber}, N.~M. and {Garcia}, P. and {Gao}, F. and {Gendron}, E. and {Genzel}, R. and {Gillessen}, S. and {Habibi}, M. and {Haubois}, X. and {Hei{\ss}el}, G. and {Henning}, T. and {Hippler}, S. and {Horrobin}, M. and {Jim{\'e}nez-Rosales}, A. and {Jochum}, L. and {Jocou}, L. and {Kaufer}, A. and {Kervella}, P. and {Lacour}, S. and {Lapeyr{\`e}re}, V. and {Le Bouquin}, J. -B. and {L{\'e}na}, P. and {Lutz}, D. and {Nowak}, M. and {Ott}, T. and {Paumard}, T. and {Perraut}, K. and {Perrin}, G. and {Pfuhl}, O. and {Rabien}, S. and {Rodr{\'\i}guez-Coira}, G. and {Shangguan}, J. and {Shimizu}, T. and {Scheithauer}, S. and {Stadler}, J. and {Straub}, O. and {Straubmeier}, C. and {Sturm}, E. and {Tacconi}, L.~J. and {Vincent}, F. and {von Fellenberg}, S. and {Waisberg}, I. and {Widmann}, F. and {Wieprecht}, E. and {Wiezorrek}, E. and {Woillez}, J. and {Yazici}, S. and {Young}, A. and {Zins}, G.},
        title = "{Improved GRAVITY astrometric accuracy from modeling optical aberrations}",
      journal = {\aap},
         year = 2021,
        month = mar,
       volume = {647},
          eid = {A59},
        pages = {A59},
          doi = {10.1051/0004-6361/202040208},
archivePrefix = {arXiv},
       eprint = {2101.12098},
 primaryClass = {astro-ph.GA},
       adsurl = {https://ui.adsabs.harvard.edu/abs/2021A&A...647A..59G}
}

@ARTICLE{2024A&A...692A.242G,
       author = {{GRAVITY Collaboration} and {Abd El Dayem}, K. and {Abuter}, R. and {Aimar}, N. and {Amaro Seoane}, P. and {Amorim}, A. and {Beck}, J. and {Berger}, J.~P. and {Bonnet}, H. and {Bourdarot}, G. and {Brandner}, W. and {Cardoso}, V. and {Capuzzo Dolcetta}, R. and {Cl{\'e}net}, Y. and {Davies}, R. and {de Zeeuw}, P.~T. and {Drescher}, A. and {Eckart}, A. and {Eisenhauer}, F. and {Feuchtgruber}, H. and {Finger}, G. and {F{\"o}rster Schreiber}, N.~M. and {Foschi}, A. and {Gao}, F. and {Garcia}, P. and {Gendron}, E. and {Genzel}, R. and {Gillessen}, S. and {Hartl}, M. and {Haubois}, X. and {Haussmann}, F. and {Hei{\ss}el}, G. and {Henning}, T. and {Hippler}, S. and {Horrobin}, M. and {Jochum}, L. and {Jocou}, L. and {Kaufer}, A. and {Kervella}, P. and {Lacour}, S. and {Lapeyr{\`e}re}, V. and {Le Bouquin}, J. -B. and {L{\'e}na}, P. and {Lutz}, D. and {Mang}, F. and {More}, N. and {Ott}, T. and {Paumard}, T. and {Perraut}, K. and {Perrin}, G. and {Pfuhl}, O. and {Rabien}, S. and {Ribeiro}, D.~C. and {Sadun Bordoni}, M. and {Scheithauer}, S. and {Shangguan}, J. and {Shimizu}, T. and {Stadler}, J. and {Straub}, O. and {Straubmeier}, C. and {Sturm}, E. and {Tacconi}, L.~J. and {Urso}, I. and {Vincent}, F. and {von Fellenberg}, S.~D. and {Widmann}, F. and {Wieprecht}, E. and {Woillez}, J. and {Zhang}, F.},
        title = "{Improving constraints on the extended mass distribution in the Galactic center with stellar orbits}",
      journal = {\aap},
         year = 2024,
        month = dec,
       volume = {692},
          eid = {A242},
        pages = {A242},
          doi = {10.1051/0004-6361/202452274},
       adsurl = {https://ui.adsabs.harvard.edu/abs/2024A&A...692A.242G}
}

@ARTICLE{2023A&A...672A..63G,
       author = {{GRAVITY Collaboration} and {Straub}, O. and {Baub{\"o}ck}, M. and {Abuter}, R. and {Aimar}, N. and {Amaro Seoane}, P. and {Amorim}, A. and {Berger}, J.~P. and {Bonnet}, H. and {Bourdarot}, G. and {Brandner}, W. and {Cardoso}, V. and {Cl{\'e}net}, Y. and {Dallilar}, Y. and {Davies}, R. and {de Zeeuw}, P.~T. and {Dexter}, J. and {Drescher}, A. and {Eisenhauer}, F. and {F{\"o}rster Schreiber}, N.~M. and {Foschi}, A. and {Garcia}, P. and {Gao}, F. and {Gendron}, E. and {Genzel}, R. and {Gillessen}, S. and {Habibi}, M. and {Haubois}, X. and {Hei{\ss}el}, G. and {Henning}, T. and {Hippler}, S. and {Horrobin}, M. and {Jochum}, L. and {Jocou}, L. and {Kaufer}, A. and {Kervella}, P. and {Lacour}, S. and {Lapeyr{\`e}re}, V. and {Le Bouquin}, J. -B. and {L{\'e}na}, P. and {Lutz}, D. and {Ott}, T. and {Paumard}, T. and {Perraut}, K. and {Perrin}, G. and {Pfuhl}, O. and {Rabien}, S. and {Ribeiro}, D.~C. and {Sadun Bordoni}, M. and {Scheithauer}, S. and {Shangguan}, J. and {Shimizu}, T. and {Stadler}, J. and {Straubmeier}, C. and {Sturm}, E. and {Tacconi}, L.~J. and {Vincent}, F. and {von Fellenberg}, S. and {Widmann}, F. and {Wieprecht}, E. and {Wiezorrek}, E. and {Woillez}, J. and {Yazici}, S.},
        title = "{Where intermediate-mass black holes could hide in the Galactic Centre. A full parameter study with the S2 orbit}",
      journal = {\aap},
         year = 2023,
        month = apr,
       volume = {672},
          eid = {A63},
        pages = {A63},
          doi = {10.1051/0004-6361/202245132},
archivePrefix = {arXiv},
       eprint = {2303.04067},
 primaryClass = {astro-ph.GA},
       adsurl = {https://ui.adsabs.harvard.edu/abs/2023A&A...672A..63G}
}

@ARTICLE{2019Sci...365..664D,
       author = {{Do}, Tuan and {Hees}, Aurelien and {Ghez}, Andrea and {Martinez}, Gregory D. and {Chu}, Devin S. and {Jia}, Siyao and {Sakai}, Shoko and {Lu}, Jessica R. and {Gautam}, Abhimat K. and {O'Neil}, Kelly Kosmo and {Becklin}, Eric E. and {Morris}, Mark R. and {Matthews}, Keith and {Nishiyama}, Shogo and {Campbell}, Randy and {Chappell}, Samantha and {Chen}, Zhuo and {Ciurlo}, Anna and {Dehghanfar}, Arezu and {Gallego-Cano}, Eulalia and {Kerzendorf}, Wolfgang E. and {Lyke}, James E. and {Naoz}, Smadar and {Saida}, Hiromi and {Sch{\"o}del}, Rainer and {Takahashi}, Masaaki and {Takamori}, Yohsuke and {Witzel}, Gunther and {Wizinowich}, Peter},
        title = "{Relativistic redshift of the star S0-2 orbiting the Galactic Center supermassive black hole}",
      journal = {Science},
         year = 2019,
        month = aug,
       volume = {365},
       number = {6454},
        pages = {664-668},
          doi = {10.1126/science.aav8137},
archivePrefix = {arXiv},
       eprint = {1907.10731},
 primaryClass = {astro-ph.GA},
       adsurl = {https://ui.adsabs.harvard.edu/abs/2019Sci...365..664D}
}

@ARTICLE{2003ApJ...599.1129Y,
       author = {{Yu}, Qingjuan and {Tremaine}, Scott},
        title = "{Ejection of Hypervelocity Stars by the (Binary) Black Hole in the Galactic Center}",
      journal = {\apj},
         year = 2003,
        month = dec,
       volume = {599},
       number = {2},
        pages = {1129-1138},
          doi = {10.1086/379546},
archivePrefix = {arXiv},
       eprint = {astro-ph/0309084},
 primaryClass = {astro-ph},
       adsurl = {https://ui.adsabs.harvard.edu/abs/2003ApJ...599.1129Y}
}

@ARTICLE{2012A&A...541A.151G,
       author = {{Giuppone}, C.~A. and {Morais}, M.~H.~M. and {Bou{\'e}}, G. and {Correia}, A.~C.~M.},
        title = "{Dynamical analysis and constraints for the HD 196885 system}",
      journal = {\aap},
         year = 2012,
        month = may,
       volume = {541},
          eid = {A151},
        pages = {A151},
          doi = {10.1051/0004-6361/201118356},
archivePrefix = {arXiv},
       eprint = {1203.5249},
 primaryClass = {astro-ph.EP},
       adsurl = {https://ui.adsabs.harvard.edu/abs/2012A&A...541A.151G}
}

@ARTICLE{2007ApJ...669.1298F,
       author = {{Fabrycky}, Daniel and {Tremaine}, Scott},
        title = "{Shrinking Binary and Planetary Orbits by Kozai Cycles with Tidal Friction}",
      journal = {\apj},
         year = 2007,
        month = nov,
       volume = {669},
       number = {2},
        pages = {1298-1315},
          doi = {10.1086/521702},
archivePrefix = {arXiv},
       eprint = {0705.4285},
 primaryClass = {astro-ph},
       adsurl = {https://ui.adsabs.harvard.edu/abs/2007ApJ...669.1298F}
}

@ARTICLE{2017ApJ...847..120H,
       author = {{Habibi}, M. and {Gillessen}, S. and {Martins}, F. and {Eisenhauer}, F. and {Plewa}, P.~M. and {Pfuhl}, O. and {George}, E. and {Dexter}, J. and {Waisberg}, I. and {Ott}, T. and {von Fellenberg}, S. and {Baub{\"o}ck}, M. and {Jimenez-Rosales}, A. and {Genzel}, R.},
        title = "{Twelve Years of Spectroscopic Monitoring in the Galactic Center: The Closest Look at S-stars near the Black Hole}",
      journal = {\apj},
         year = 2017,
        month = oct,
       volume = {847},
       number = {2},
          eid = {120},
        pages = {120},
          doi = {10.3847/1538-4357/aa876f},
archivePrefix = {arXiv},
       eprint = {1708.06353},
 primaryClass = {astro-ph.SR},
       adsurl = {https://ui.adsabs.harvard.edu/abs/2017ApJ...847..120H}
}

@ARTICLE{2023MNRAS.519.3281B,
       author = {{Boekholt}, Tjarda C.~N. and {Vaillant}, Timoth{\'e}e and {Correia}, Alexandre C.~M.},
        title = "{Reversible time-step adaptation for the integration of few-body systems}",
      journal = {\mnras},
         year = 2023,
        month = mar,
       volume = {519},
       number = {3},
        pages = {3281-3291},
          doi = {10.1093/mnras/stac3777},
archivePrefix = {arXiv},
       eprint = {2212.09745},
 primaryClass = {astro-ph.IM},
       adsurl = {https://ui.adsabs.harvard.edu/abs/2023MNRAS.519.3281B}
}

@ARTICLE{1988Natur.331..687H,
       author = {{Hills}, J.~G.},
        title = "{Hyper-velocity and tidal stars from binaries disrupted by a massive Galactic black hole}",
      journal = {\nat},
         year = 1988,
        month = feb,
       volume = {331},
       number = {6158},
        pages = {687-689},
          doi = {10.1038/331687a0},
       adsurl = {https://ui.adsabs.harvard.edu/abs/1988Natur.331..687H}
}

@ARTICLE{2008ApJ...683L.151L,
       author = {{L{\"o}ckmann}, Ulf and {Baumgardt}, Holger and {Kroupa}, Pavel},
        title = "{Origin of the S Stars in the Galactic Center}",
      journal = {\apjl},
         year = 2008,
        month = aug,
       volume = {683},
       number = {2},
        pages = {L151},
          doi = {10.1086/591734},
archivePrefix = {arXiv},
       eprint = {0807.2239},
 primaryClass = {astro-ph},
       adsurl = {https://ui.adsabs.harvard.edu/abs/2008ApJ...683L.151L}
}

@ARTICLE{2018A&A...615L..15G,
       author = {{GRAVITY Collaboration} and {Abuter}, R. and {Amorim}, A. and {Anugu}, N. and {Baub{\"o}ck}, M. and {Benisty}, M. and {Berger}, J.~P. and {Blind}, N. and {Bonnet}, H. and {Brandner}, W. and {Buron}, A. and {Collin}, C. and {Chapron}, F. and {Cl{\'e}net}, Y. and {Coud{\'e} Du Foresto}, V. and {de Zeeuw}, P.~T. and {Deen}, C. and {Delplancke-Str{\"o}bele}, F. and {Dembet}, R. and {Dexter}, J. and {Duvert}, G. and {Eckart}, A. and {Eisenhauer}, F. and {Finger}, G. and {F{\"o}rster Schreiber}, N.~M. and {F{\'e}dou}, P. and {Garcia}, P. and {Garcia Lopez}, R. and {Gao}, F. and {Gendron}, E. and {Genzel}, R. and {Gillessen}, S. and {Gordo}, P. and {Habibi}, M. and {Haubois}, X. and {Haug}, M. and {Hau{\ss}mann}, F. and {Henning}, Th. and {Hippler}, S. and {Horrobin}, M. and {Hubert}, Z. and {Hubin}, N. and {Jimenez Rosales}, A. and {Jochum}, L. and {Jocou}, K. and {Kaufer}, A. and {Kellner}, S. and {Kendrew}, S. and {Kervella}, P. and {Kok}, Y. and {Kulas}, M. and {Lacour}, S. and {Lapeyr{\`e}re}, V. and {Lazareff}, B. and {Le Bouquin}, J.-B. and {L{\'e}na}, P. and {Lippa}, M. and {Lenzen}, R. and {M{\'e}rand}, A. and {M{\"u}ler}, E. and {Neumann}, U. and {Ott}, T. and {Palanca}, L. and {Paumard}, T. and {Pasquini}, L. and {Perraut}, K. and {Perrin}, G. and {Pfuhl}, O. and {Plewa}, P.~M. and {Rabien}, S. and {Ram{\'\i}rez}, A. and {Ramos}, J. and {Rau}, C. and {Rodr{\'\i}guez-Coira}, G. and {Rohloff}, R.-R. and {Rousset}, G. and {Sanchez-Bermudez}, J. and {Scheithauer}, S. and {Sch{\"o}ller}, M. and {Schuler}, N. and {Spyromilio}, J. and {Straub}, O. and {Straubmeier}, C. and {Sturm}, E. and {Tacconi}, L.~J. and {Tristram}, K.~R.~W. and {Vincent}, F. and {von Fellenberg}, S. and {Wank}, I. and {Waisberg}, I. and {Widmann}, F. and {Wieprecht}, E. and {Wiest}, M. and {Wiezorrek}, E. and {Woillez}, J. and {Yazici}, S. and {Ziegler}, D. and {Zins}, G.},
        title = "{Detection of the gravitational redshift in the orbit of the star S2 near the Galactic centre massive black hole}",
      journal = {\aap},
         year = 2018,
        month = jul,
       volume = {615},
          eid = {L15},
        pages = {L15},
          doi = {10.1051/0004-6361/201833718},
archivePrefix = {arXiv},
       eprint = {1807.09409},
 primaryClass = {astro-ph.GA},
       adsurl = {https://ui.adsabs.harvard.edu/abs/2018A&A...615L..15G}
}

@ARTICLE{2023ApJ...948...94C,
       author = {{Chu}, Devin S. and {Do}, Tuan and {Ghez}, Andrea and {Gautam}, Abhimat K. and {Ciurlo}, Anna and {O'neil}, Kelly Kosmo and {Hosek}, Matthew W. and {Hees}, Aur{\'e}lien and {Naoz}, Smadar and {Sakai}, Shoko and {Lu}, Jessica R. and {Chen}, Zhuo and {Bentley}, Rory O. and {Becklin}, Eric E. and {Matthews}, Keith},
        title = "{Evidence of a Decreased Binary Fraction for Massive Stars within 20 milliparsecs of the Supermassive Black Hole at the Galactic Center}",
      journal = {\apj},
         year = 2023,
        month = may,
       volume = {948},
       number = {2},
          eid = {94},
        pages = {94},
          doi = {10.3847/1538-4357/acc93e},
archivePrefix = {arXiv},
       eprint = {2303.16977},
 primaryClass = {astro-ph.GA},
       adsurl = {https://ui.adsabs.harvard.edu/abs/2023ApJ...948...94C}
}

@ARTICLE{2012Sci...337..444S,
       author = {{Sana}, H. and {de Mink}, S.~E. and {de Koter}, A. and {Langer}, N. and {Evans}, C.~J. and {Gieles}, M. and {Gosset}, E. and {Izzard}, R.~G. and {Le Bouquin}, J. -B. and {Schneider}, F.~R.~N.},
        title = "{Binary Interaction Dominates the Evolution of Massive Stars}",
      journal = {Science},
         year = 2012,
        month = jul,
       volume = {337},
       number = {6093},
        pages = {444},
          doi = {10.1126/science.1223344},
archivePrefix = {arXiv},
       eprint = {1207.6397},
 primaryClass = {astro-ph.SR},
       adsurl = {https://ui.adsabs.harvard.edu/abs/2012Sci...337..444S}
}

@ARTICLE{2023MNRAS.522.2885B,
       author = {{Boekholt}, Tjarda C.~N. and {Correia}, Alexandre C.~M.},
        title = "{A direct N-body integrator for modelling the chaotic, tidal dynamics of multibody extrasolar systems: TIDYMESS}",
      journal = {\mnras},
         year = 2023,
        month = jun,
       volume = {522},
       number = {2},
        pages = {2885-2900},
          doi = {10.1093/mnras/stad1133},
archivePrefix = {arXiv},
       eprint = {2209.03955},
 primaryClass = {astro-ph.IM},
       adsurl = {https://ui.adsabs.harvard.edu/abs/2023MNRAS.522.2885B}
}

@ARTICLE{2022A&A...657L..12G,
       author = {{GRAVITY Collaboration} and {Abuter}, R. and {Aimar}, N. and {Amorim}, A. and {Ball}, J. and {Baub{\"o}ck}, M. and {Berger}, J.~P. and {Bonnet}, H. and {Bourdarot}, G. and {Brandner}, W. and {Cardoso}, V. and {Cl{\'e}net}, Y. and {Dallilar}, Y. and {Davies}, R. and {de Zeeuw}, P.~T. and {Dexter}, J. and {Drescher}, A. and {Eisenhauer}, F. and {F{\"o}rster Schreiber}, N.~M. and {Foschi}, A. and {Garcia}, P. and {Gao}, F. and {Gendron}, E. and {Genzel}, R. and {Gillessen}, S. and {Habibi}, M. and {Haubois}, X. and {Hei{\ss}el}, G. and {Henning}, T. and {Hippler}, S. and {Horrobin}, M. and {Jochum}, L. and {Jocou}, L. and {Kaufer}, A. and {Kervella}, P. and {Lacour}, S. and {Lapeyr{\`e}re}, V. and {Le Bouquin}, J.-B. and {L{\'e}na}, P. and {Lutz}, D. and {Ott}, T. and {Paumard}, T. and {Perraut}, K. and {Perrin}, G. and {Pfuhl}, O. and {Rabien}, S. and {Shangguan}, J. and {Shimizu}, T. and {Scheithauer}, S. and {Stadler}, J. and {Stephens}, A.~W. and {Straub}, O. and {Straubmeier}, C. and {Sturm}, E. and {Tacconi}, L.~J. and {Tristram}, K.~R.~W. and {Vincent}, F. and {von Fellenberg}, S. and {Widmann}, F. and {Wieprecht}, E. and {Wiezorrek}, E. and {Woillez}, J. and {Yazici}, S. and {Young}, A.},
        title = "{Mass distribution in the Galactic Center based on interferometric astrometry of multiple stellar orbits}",
      journal = {\aap},
         year = 2022,
        month = jan,
       volume = {657},
          eid = {L12},
        pages = {L12},
          doi = {10.1051/0004-6361/202142465},
archivePrefix = {arXiv},
       eprint = {2112.07478},
 primaryClass = {astro-ph.GA},
       adsurl = {https://ui.adsabs.harvard.edu/abs/2022A&A...657L..12G}
}

@ARTICLE{2023MNRAS.524.1075F,
       author = {{GRAVITY Collaboration} and {Foschi}, A. and {Abuter}, R. and {Aimar}, N. and {Amaro Seoane}, P. and {Amorim}, A. and {Baub{\"o}ck}, M. and {Berger}, J.~P. and {Bonnet}, H. and {Bourdarot}, G. and {Brandner}, W. and {Cardoso}, V. and {Cl{\'e}net}, Y. and {Dallilar}, Y. and {Davies}, R. and {de Zeeuw}, P.~T. and {Defr{\`e}re}, D. and {Dexter}, J. and {Drescher}, A. and {Eckart}, A. and {Eisenhauer}, F. and {Ferreira}, M.~C. and {F{\"o}rster Schreiber}, N.~M. and {Garcia}, P.~J.~V. and {Gao}, F. and {Gendron}, E. and {Genzel}, R. and {Gillessen}, S. and {Gomes}, T. and {Habibi}, M. and {Haubois}, X. and {Hei{\ss}el}, G. and {Henning}, T. and {Hippler}, S. and {H{\"o}nig}, S.~F. and {Horrobin}, M. and {Jochum}, L. and {Jocou}, L. and {Kaufer}, A. and {Kervella}, P. and {Kreidberg}, L. and {Lacour}, S. and {Lapeyr{\`e}re}, V. and {Le Bouquin}, J. -B. and {L{\'e}na}, P. and {Lutz}, D. and {Millour}, F. and {Ott}, T. and {Paumard}, T. and {Perraut}, K. and {Perrin}, G. and {Pfuhl}, O. and {Rabien}, S. and {Ribeiro}, D.~C. and {Sadun Bordoni}, M. and {Scheithauer}, S. and {Shangguan}, J. and {Shimizu}, T. and {Stadler}, J. and {Straub}, O. and {Straubmeier}, C. and {Sturm}, E. and {Sykes}, C. and {Tacconi}, L.~J. and {Vincent}, F. and {von Fellenberg}, S. and {Widmann}, F. and {Wieprecht}, E. and {Wiezorrek}, E. and {Woillez}, J. and {GRAVITY Collaboration}},
        title = "{Using the motion of S2 to constrain scalar clouds around Sgr A*}",
      journal = {\mnras},
         year = 2023,
        month = sep,
       volume = {524},
       number = {1},
        pages = {1075-1086},
          doi = {10.1093/mnras/stad1939},
archivePrefix = {arXiv},
       eprint = {2306.17215},
 primaryClass = {astro-ph.GA},
       adsurl = {https://ui.adsabs.harvard.edu/abs/2023MNRAS.524.1075F}
}

@ARTICLE{2010ApJ...713...90A,
       author = {{Antonini}, Fabio and {Faber}, Joshua and {Gualandris}, Alessia and {Merritt}, David},
        title = "{Tidal Breakup of Binary Stars at the Galactic Center and Its Consequences}",
      journal = {\apj},
         year = 2010,
        month = apr,
       volume = {713},
       number = {1},
        pages = {90-104},
          doi = {10.1088/0004-637X/713/1/90},
archivePrefix = {arXiv},
       eprint = {0909.1959},
 primaryClass = {astro-ph.GA},
       adsurl = {https://ui.adsabs.harvard.edu/abs/2010ApJ...713...90A}
}

@ARTICLE{2016MNRAS.460.3494S,
       author = {{Stephan}, Alexander P. and {Naoz}, Smadar and {Ghez}, Andrea M. and {Witzel}, Gunther and {Sitarski}, Breann N. and {Do}, Tuan and {Kocsis}, Bence},
        title = "{Merging binaries in the Galactic Center: the eccentric Kozai-Lidov mechanism with stellar evolution}",
      journal = {\mnras},
         year = 2016,
        month = aug,
       volume = {460},
       number = {4},
        pages = {3494-3504},
          doi = {10.1093/mnras/stw1220},
archivePrefix = {arXiv},
       eprint = {1603.02709},
 primaryClass = {astro-ph.SR},
       adsurl = {https://ui.adsabs.harvard.edu/abs/2016MNRAS.460.3494S}
}

@ARTICLE{2016ARA&A..54..441N,
       author = {{Naoz}, Smadar},
        title = "{The Eccentric Kozai-Lidov Effect and Its Applications}",
      journal = {\araa},
         year = 2016,
        month = sep,
       volume = {54},
        pages = {441-489},
          doi = {10.1146/annurev-astro-081915-023315},
archivePrefix = {arXiv},
       eprint = {1601.07175},
 primaryClass = {astro-ph.EP},
       adsurl = {https://ui.adsabs.harvard.edu/abs/2016ARA&A..54..441N}
}

@ARTICLE{1962P&SS....9..719L,
       author = {{Lidov}, M.~L.},
        title = "{The evolution of orbits of artificial satellites of planets under the action of gravitational perturbations of external bodies}",
      journal = {\planss},
         year = 1962,
        month = oct,
       volume = {9},
       number = {10},
        pages = {719-759},
          doi = {10.1016/0032-0633(62)90129-0},
       adsurl = {https://ui.adsabs.harvard.edu/abs/1962P&SS....9..719L}
}

@ARTICLE{1962AJ.....67..591K,
       author = {{Kozai}, Yoshihide},
        title = "{Secular perturbations of asteroids with high inclination and eccentricity}",
      journal = {\aj},
         year = 1962,
        month = nov,
       volume = {67},
        pages = {591-598},
          doi = {10.1086/108790},
       adsurl = {https://ui.adsabs.harvard.edu/abs/1962AJ.....67..591K}
}

@ARTICLE{2018MNRAS.479.4749B,
       author = {{Bataille}, M. and {Libert}, A. -S. and {Correia}, A.~C.~M.},
        title = "{Dynamical evolution of triple-star systems by Lidov-Kozai cycles and tidal friction}",
      journal = {\mnras},
         year = 2018,
        month = oct,
       volume = {479},
       number = {4},
        pages = {4749-4759},
          doi = {10.1093/mnras/sty1758},
archivePrefix = {arXiv},
       eprint = {1807.05901},
 primaryClass = {astro-ph.EP},
       adsurl = {https://ui.adsabs.harvard.edu/abs/2018MNRAS.479.4749B}
}

@BOOK{Jeans_1919B,
       author = {{Jeans}, James Hopwood},
        title = "{Problems of cosmogony and stellar dynamics}",
         year = 1919,
       adsurl = {https://ui.adsabs.harvard.edu/abs/1919pcsd.book.....J}
}

@ARTICLE{Roche_1849,
       author = {{Roche}, \'Edouard},
        title = "{La figure d'une masse fluide soumise \`a l'attraction d'un point \'eloign\'e}",
      journal = {Acad\'emie des Sciences de Montpellier: M\'emoires de la Section des Sciences},
         year = 1849,
       volume = {1},
        pages = {243-262},
       adsurl = {https://books.google.com/books?id=UmoVAAAAQAAJ&pg=PA243}
}

@ARTICLE{2018ApJ...854...12C,
       author = {{Chu}, Devin S. and {Do}, Tuan and {Hees}, Aurelien and {Ghez}, Andrea and {Naoz}, Smadar and {Witzel}, Gunther and {Sakai}, Shoko and {Chappell}, Samantha and {Gautam}, Abhimat K. and {Lu}, Jessica R. and {Matthews}, Keith},
        title = "{Investigating the Binarity of S0-2: Implications for Its Origins and Robustness as a Probe of the Laws of Gravity around a Supermassive Black Hole}",
      journal = {\apj},
         year = 2018,
        month = feb,
       volume = {854},
       number = {1},
          eid = {12},
        pages = {12},
          doi = {10.3847/1538-4357/aaa3eb},
archivePrefix = {arXiv},
       eprint = {1709.04890},
 primaryClass = {astro-ph.SR},
       adsurl = {https://ui.adsabs.harvard.edu/abs/2018ApJ...854...12C}
}

@ARTICLE{2012MNRAS.424...52M,
       author = {{Morais}, M.~H.~M. and {Giuppone}, C.~A.},
        title = "{Stability of prograde and retrograde planets in circular binary systems}",
      journal = {\mnras},
         year = 2012,
        month = jul,
       volume = {424},
       number = {1},
        pages = {52-64},
          doi = {10.1111/j.1365-2966.2012.21151.x},
archivePrefix = {arXiv},
       eprint = {1204.4718},
 primaryClass = {astro-ph.EP},
       adsurl = {https://ui.adsabs.harvard.edu/abs/2012MNRAS.424...52M}
}

@ARTICLE{2010MNRAS.401.1189F,
       author = {{Farago}, F. and {Laskar}, J.},
        title = "{High-inclination orbits in the secular quadrupolar three-body problem}",
      journal = {\mnras},
         year = 2010,
        month = jan,
       volume = {401},
       number = {2},
        pages = {1189-1198},
          doi = {10.1111/j.1365-2966.2009.15711.x},
archivePrefix = {arXiv},
       eprint = {0909.2287},
 primaryClass = {astro-ph.EP},
       adsurl = {https://ui.adsabs.harvard.edu/abs/2010MNRAS.401.1189F}
}

@INCOLLECTION{2010exop.book...15M,
       author = {{Murray}, C.~D. and {Correia}, A.~C.~M.},
        title = "{Keplerian Orbits and Dynamics of Exoplanets}",
    booktitle = {Exoplanets},
         year = 2010,
       editor = {{Seager}, S.},
        pages = {15-23},
          doi = {10.48550/arXiv.1009.1738},
       adsurl = {https://ui.adsabs.harvard.edu/abs/2010exop.book...15M}
}

@ARTICLE{2019A&A...625L..10G,
       author = {{GRAVITY Collaboration} and {Abuter}, R. and {Amorim}, A. and {Baub{\"o}ck}, M. and {Berger}, J.~P. and {Bonnet}, H. and {Brandner}, W. and {Cl{\'e}net}, Y. and {Coud{\'e} Du Foresto}, V. and {de Zeeuw}, P.~T. and {Dexter}, J. and {Duvert}, G. and {Eckart}, A. and {Eisenhauer}, F. and {F{\"o}rster Schreiber}, N.~M. and {Garcia}, P. and {Gao}, F. and {Gendron}, E. and {Genzel}, R. and {Gerhard}, O. and {Gillessen}, S. and {Habibi}, M. and {Haubois}, X. and {Henning}, T. and {Hippler}, S. and {Horrobin}, M. and {Jim{\'e}nez-Rosales}, A. and {Jocou}, L. and {Kervella}, P. and {Lacour}, S. and {Lapeyr{\`e}re}, V. and {Le Bouquin}, J.-B. and {L{\'e}na}, P. and {Ott}, T. and {Paumard}, T. and {Perraut}, K. and {Perrin}, G. and {Pfuhl}, O. and {Rabien}, S. and {Rodriguez Coira}, G. and {Rousset}, G. and {Scheithauer}, S. and {Sternberg}, A. and {Straub}, O. and {Straubmeier}, C. and {Sturm}, E. and {Tacconi}, L.~J. and {Vincent}, F. and {von Fellenberg}, S. and {Waisberg}, I. and {Widmann}, F. and {Wieprecht}, E. and {Wiezorrek}, E. and {Woillez}, J. and {Yazici}, S.},
        title = "{A geometric distance measurement to the Galactic center black hole with 0.3\% uncertainty}",
      journal = {\aap},
         year = 2019,
        month = may,
       volume = {625},
          eid = {L10},
        pages = {L10},
          doi = {10.1051/0004-6361/201935656},
archivePrefix = {arXiv},
       eprint = {1904.05721},
 primaryClass = {astro-ph.GA},
       adsurl = {https://ui.adsabs.harvard.edu/abs/2019A&A...625L..10G}
}

@ARTICLE{2020A&A...636L...5G,
       author = {{GRAVITY Collaboration} and {Abuter}, R. and {Amorim}, A. and {Baub{\"o}ck}, M. and {Berger}, J.~P. and {Bonnet}, H. and {Brandner}, W. and {Cardoso}, V. and {Cl{\'e}net}, Y. and {de Zeeuw}, P.~T. and {Dexter}, J. and {Eckart}, A. and {Eisenhauer}, F. and {F{\"o}rster Schreiber}, N.~M. and {Garcia}, P. and {Gao}, F. and {Gendron}, E. and {Genzel}, R. and {Gillessen}, S. and {Habibi}, M. and {Haubois}, X. and {Henning}, T. and {Hippler}, S. and {Horrobin}, M. and {Jim{\'e}nez-Rosales}, A. and {Jochum}, L. and {Jocou}, L. and {Kaufer}, A. and {Kervella}, P. and {Lacour}, S. and {Lapeyr{\`e}re}, V. and {Le Bouquin}, J.-B. and {L{\'e}na}, P. and {Nowak}, M. and {Ott}, T. and {Paumard}, T. and {Perraut}, K. and {Perrin}, G. and {Pfuhl}, O. and {Rodr{\'\i}guez-Coira}, G. and {Shangguan}, J. and {Scheithauer}, S. and {Stadler}, J. and {Straub}, O. and {Straubmeier}, C. and {Sturm}, E. and {Tacconi}, L.~J. and {Vincent}, F. and {von Fellenberg}, S. and {Waisberg}, I. and {Widmann}, F. and {Wieprecht}, E. and {Wiezorrek}, E. and {Woillez}, J. and {Yazici}, S. and {Zins}, G.},
        title = "{Detection of the Schwarzschild precession in the orbit of the star S2 near the Galactic centre massive black hole}",
      journal = {\aap},
         year = 2020,
        month = apr,
       volume = {636},
          eid = {L5},
        pages = {L5},
          doi = {10.1051/0004-6361/202037813},
archivePrefix = {arXiv},
       eprint = {2004.07187},
 primaryClass = {astro-ph.GA},
       adsurl = {https://ui.adsabs.harvard.edu/abs/2020A&A...636L...5G}
}

@ARTICLE{2019PhRvL.122j1102A,
       author = {{Amorim}, A. and {Baub{\"o}ck}, M. and {Berger}, J.~P. and {Brandner}, W. and {Cl{\'e}net}, Y. and {Coud{\'e} Du Foresto}, V. and {de Zeeuw}, P.~T. and {Dexter}, J. and {Duvert}, G. and {Ebert}, M. and {Eckart}, A. and {Eisenhauer}, F. and {F{\"o}rster Schreiber}, N.~M. and {Garcia}, P. and {Gao}, F. and {Gendron}, E. and {Genzel}, R. and {Gillessen}, S. and {Habibi}, M. and {Haubois}, X. and {Henning}, Th. and {Hippler}, S. and {Horrobin}, M. and {Hubert}, Z. and {Jim{\'e}nez Rosales}, A. and {Jocou}, L. and {Kervella}, P. and {Lacour}, S. and {Lapeyr{\`e}re}, V. and {Le Bouquin}, J. -B. and {L{\'e}na}, P. and {Ott}, T. and {Paumard}, T. and {Perraut}, K. and {Perrin}, G. and {Pfuhl}, O. and {Rabien}, S. and {Rodr{\'\i}guez-Coira}, G. and {Rousset}, G. and {Scheithauer}, S. and {Sternberg}, A. and {Straub}, O. and {Straubmeier}, C. and {Sturm}, E. and {Tacconi}, L.~J. and {Vincent}, F. and {von Fellenberg}, S. and {Waisberg}, I. and {Widmann}, F. and {Wieprecht}, E. and {Wiezorrek}, E. and {Yazici}, S. and {GRAVITY Collaboration}},
        title = "{Test of the Einstein Equivalence Principle near the Galactic Center Supermassive Black Hole}",
      journal = {\prl},
         year = 2019,
        month = mar,
       volume = {122},
       number = {10},
          eid = {101102},
        pages = {101102},
          doi = {10.1103/PhysRevLett.122.101102},
archivePrefix = {arXiv},
       eprint = {1902.04193},
 primaryClass = {astro-ph.GA},
       adsurl = {https://ui.adsabs.harvard.edu/abs/2019PhRvL.122j1102A}
}

@ARTICLE{2007MNRAS.374..515L,
       author = {{Levin}, Yuri},
        title = "{Starbursts near supermassive black holes: young stars in the Galactic Centre, and gravitational waves in LISA band}",
      journal = {\mnras},
         year = 2007,
        month = jan,
       volume = {374},
       number = {2},
        pages = {515-524},
          doi = {10.1111/j.1365-2966.2006.11155.x},
archivePrefix = {arXiv},
       eprint = {astro-ph/0603583},
 primaryClass = {astro-ph},
       adsurl = {https://ui.adsabs.harvard.edu/abs/2007MNRAS.374..515L}
}

@ARTICLE{2006MNRAS.368..221G,
       author = {{Ginsburg}, Idan and {Loeb}, Abraham},
        title = "{The fate of former companions to hypervelocity stars originating at the Galactic Centre}",
      journal = {\mnras},
         year = 2006,
        month = may,
       volume = {368},
       number = {1},
        pages = {221-225},
          doi = {10.1111/j.1365-2966.2006.10091.x},
archivePrefix = {arXiv},
       eprint = {astro-ph/0510574},
 primaryClass = {astro-ph},
       adsurl = {https://ui.adsabs.harvard.edu/abs/2006MNRAS.368..221G}
}

@ARTICLE{2005PhR...419...65A,
       author = {{Alexander}, Tal},
        title = "{Stellar processes near the massive black hole in the Galactic center [review article]}",
      journal = {\physrep},
         year = 2005,
        month = nov,
       volume = {419},
       number = {2-3},
        pages = {65-142},
          doi = {10.1016/j.physrep.2005.08.002},
archivePrefix = {arXiv},
       eprint = {astro-ph/0508106},
 primaryClass = {astro-ph},
       adsurl = {https://ui.adsabs.harvard.edu/abs/2005PhR...419...65A}
}

\appendix

\section{Data availibility}

The data used in this study to perform the simulations with the $N$-body code \texttt{TIDYMESS} \citep{2023MNRAS.522.2885B} and to generate all plots and conclusions are publicly available on Zenodo at  \hyperlink{https://doi.org/10.5281/zenodo.17712655}{DOI: 10.5281/zenodo.17712655} under the standard Zenodo license (Creative Commons Attribution 4.0 International — CC BY 4.0).

The database is optimized for use with SQL engines (e.g., PostgreSQL or MySQL) or their Python equivalents (e.g., SQLAlchemy). It provides detailed information about each binary system considered, including data for every simulation time-step (in years) up to the total simulation time if the system survives, or up to the time of disruption if it occurs. Additionally, the database includes the inertial position and velocity vectors of the SMBH, the primary, and the secondary components of the binary. It contains the binary’s orbital period, eccentricity, mass ratio, primary mass, and the cosine of the mutual inclination.

The database available on Zenodo combined with \texttt{TIDYMESS} enables a wide range of analyses. For instance, users may select any surviving binary and continue its evolution beyond the final simulation time considered in this paper by setting the flag \texttt{to\_continue = 1} and updating \texttt{t\_end} in \texttt{TIDYMESS}. In the case of mergers or disrupted binaries, this flag may also be enabled to evolve the resulting unbound or merged systems.

Finally, using the SQL engine and the database provided, one can fully reproduce every figure presented in this work and is enabled to conduct a deeper exploration of the simulations.

\section{Tidal and rotational perturbations}
\label{Appendix:Distortions}

The orbits of close-in binary systems also precess owing to tidal and rotational distortions of both stars. One way to assess whether these effects are relevant to our problem is to compare the timescales on which they act to those of the quadrupole approximation and general relativity.

The timescales for the phenomena considered in our work are given by \cite[e.g.,][]{2015MNRAS.451.1341L, 2016ARA&A..54..441N}.
\begin{align}
    &t_{\rm quad} \sim \frac{2\pi a_{\bullet}^{3}\left(1 - e_{\bullet}^{2} \right)^{3/2}\sqrt{\left(m_{\rm A} + m_{\rm B}\right)\left(1 - e^{2}\right)}}{\sqrt{G}a^{3/2}m_{\bullet}} \ ,  \label{t_quad}
    \\
    &t_{\rm GR} \sim \frac{2\pi a^{5/2}c^{2}\left(1 - e^{2}\right)}{3G^{3/2}\left(m_{\rm A} + m_{\rm B}\right)^{3/2}} \ ,  \label{t_gr}
\end{align}
where $a_{\bullet}$ and $e_{\bullet}$ are the semi-major axis and eccentricity of the inner pair around the SMBH (Table~\ref{orb_elem_S2}), while the tidal and rotational timescales may be defined as
\begin{align}
    &t_{\rm rot} \sim  \frac{2 G^{1/4}a^{7/2}m_{ i}\left(1 - e^{2} \right)^{2}}{k_{i} \Omega_{ i}^{2}(m_{i} + m_{j})^{1/2} R_{ i}^{5}} \ ,  \label{t_rot} \\
    &t_{\rm tide} \sim \frac{a^{13/2}\left(1 - e^{2} \right)^{5}m_{ i}}{G^{1/4} k_{i}m_{ j}\left(m_{ j} + m_{i}\right)\left(1 + \frac{3}{2}e^{2} + \frac{1}{8}e^{4} \right)R_{ i}^5 } \ , \label{t_tide}
\end{align}
where $G$ is the gravitational constant, $\Omega_{\rm i}$ is the rotation rate, $k_{i}$ is the second Love number, and $\{i,j\} = \{\rm A,\rm B\}$ or $\{i,j\} = \{\rm B,\rm A\}$.
The rotation rate and the Love numbers depend on the mass and age of the stars. 
Assuming an age of approximately 4~Myr, limited by catalog completeness, we adopt $\Omega_{i}$ from \citet{1965Obs....85..166M} and use the Love numbers derived from stellar evolution models by \citet{2004A&A...424..919C}.

\begin{figure}
    \centering
    \includegraphics[width=0.95\columnwidth]{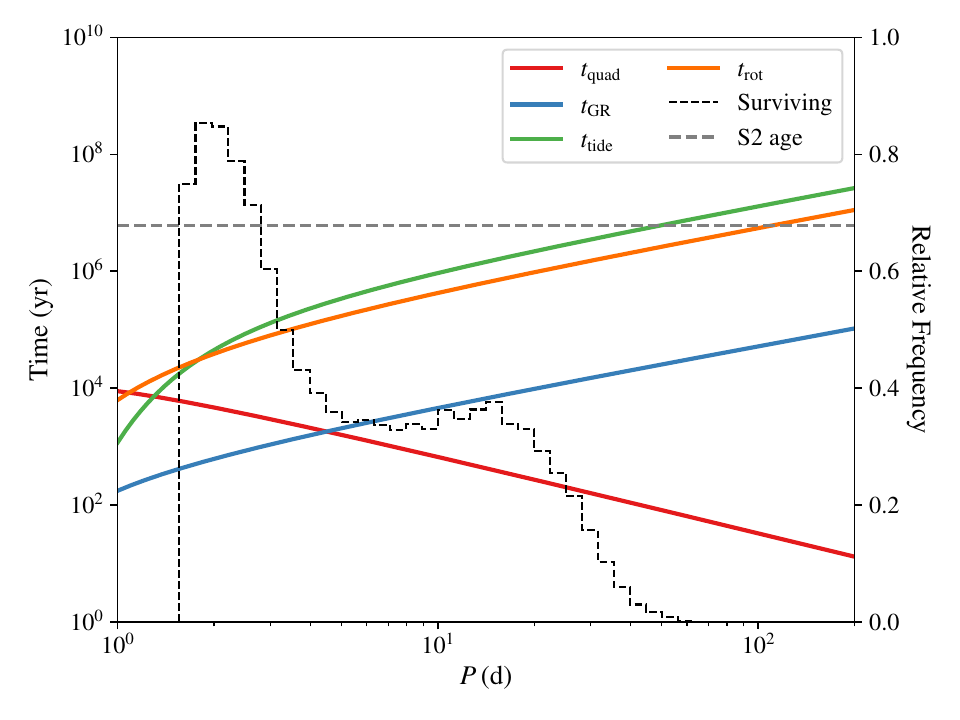}
    \caption{Characteristic evolution timescales for a typical S2-binary disturbed by the SMBH. We show the timescales for the quadrupole approximation in red (Eq.\,(\ref{t_quad})), for general relativistic precession in blue (Eq.\,(\ref{t_gr})), for rotational deformation in orange (Eq.\,(\ref{t_rot})), and for tidal deformation in green (Eq.\,(\ref{t_tide})). The black dashed line displays the histogram distribution of the surviving binaries (Fig.~\ref{P_end}\,(a)) and the grey dashed line displays the S2's age \citep{2017ApJ...847..120H}. \label{timescales_hist}}
\end{figure}

In Fig.~\ref{timescales_hist}, we show the different timescales as a function of the orbital periods for a typical S2 binary system disturbed by the SMBH. 
We observe that $t_{\rm rot}, t_{\rm tide} \ll t_{\rm GR}$, which confirms that the contribution of tidal and rotational deformations can be neglected in this problem.
Moreover, for orbital periods in the range $1.559 < P_{\rm orb} < 2.5$~d, we also have $t_{GR}< t_{\rm quad}$, and so we expect that general relativistic precession dominates the quadrupole interactions and thus suppress the Lidov-Kozai cycles. 
Indeed, by comparing with the histogram of the distribution of surviving binaries in the background, we note that the sharp decrease in the number of survivals exactly coincides with the transition of regimes ($t_{GR} \sim  t_{\rm quad}$).

\section{Analytical predictions}
\label{Appendix:AnalyticalPredictions}

\begin{figure*}
\centering
\includegraphics[width=0.9\columnwidth]{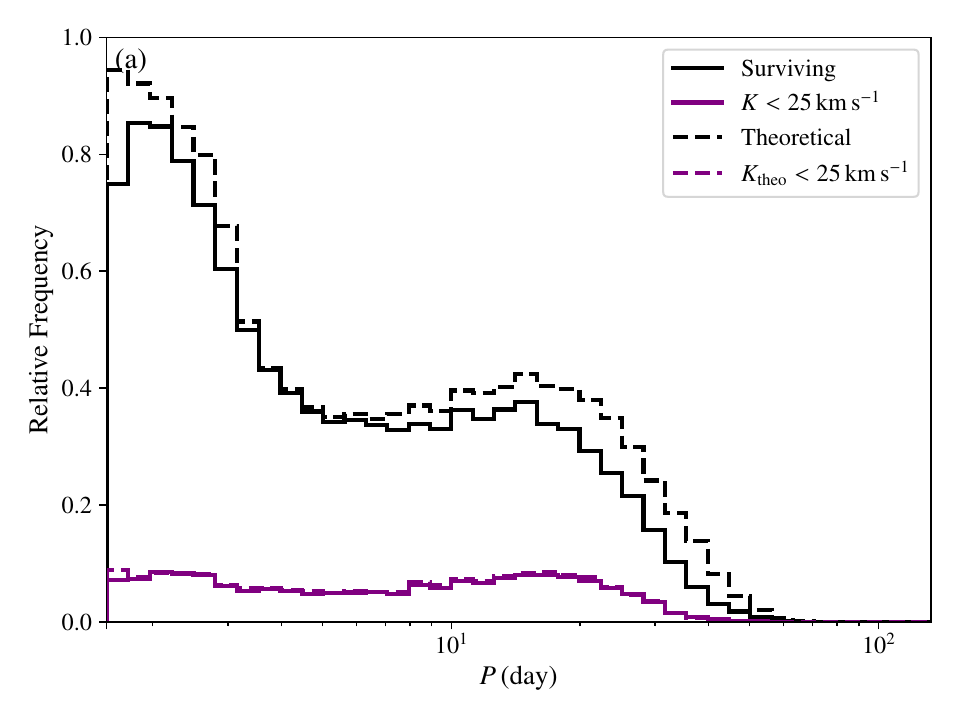} 
\includegraphics[width=0.9\columnwidth]{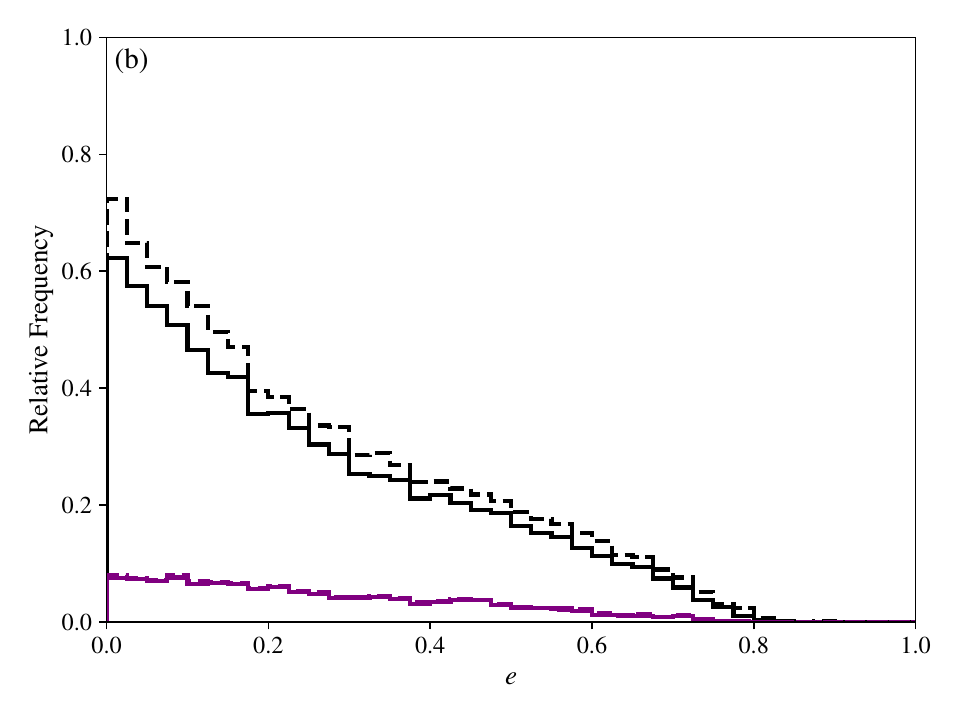} \\
\includegraphics[width=0.9\columnwidth]{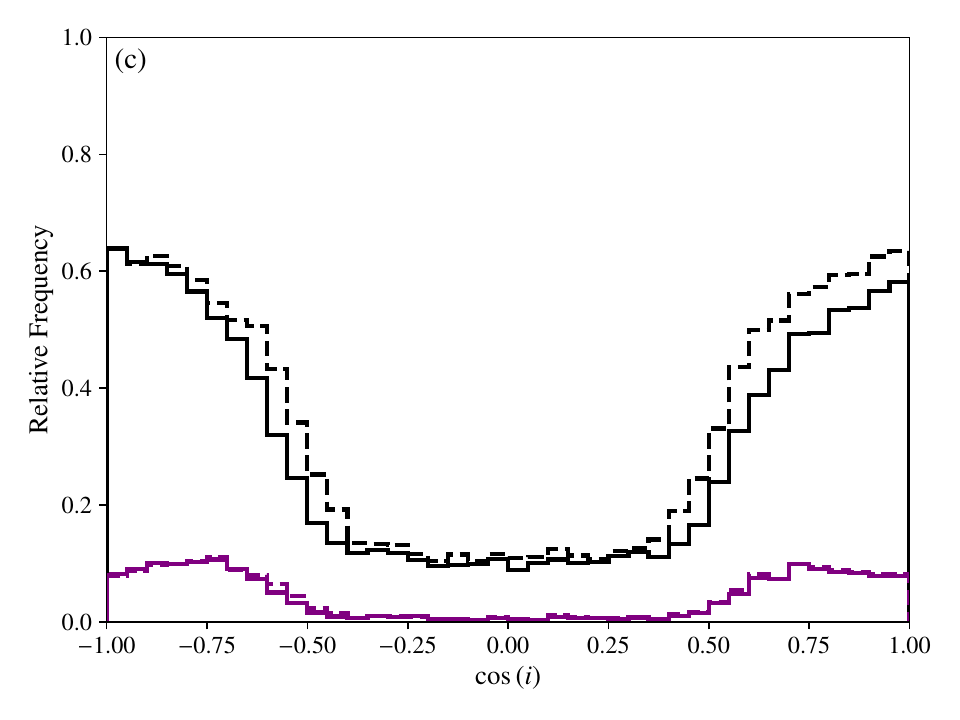} 
\includegraphics[width=0.9\columnwidth]{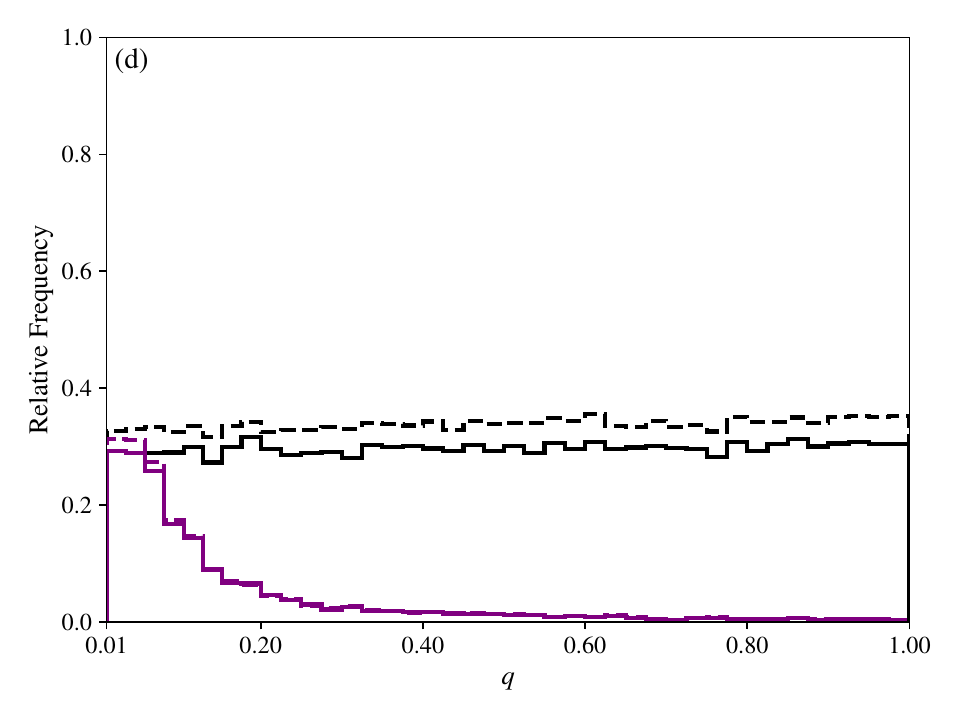}
\caption{Histograms of the relative frequency distribution of the main orbital parameters of surviving binaries. The dashed lines indicate the analytical prediction of surviving binaries based on simple criteria (Eqs.~(\ref{rochesurvive}) and (\ref{hillsurvive})), while the solid lines depict the results from the three-body numerical simulations (Fig.~\ref{P_end}). We show the orbital period (a), eccentricity (b), mutual inclination (c), and mass ratio (d). \label{Theo_vs_sim}}
\end{figure*}

In this work, we argue that Lidov-Kozai oscillations drive mergers and that  breakups occur when the system exceeds the orbit's Hill radius. Indeed, one can use some simple analytical estimations to obtain back-of-the-envelope predictions for the system's evolution based solely on the initial conditions.

To access the initial conditions destroyed by mergers through Lidov-Kozai oscillations, we consider the secular quadrupole-level Hamiltonian with general relativity corrections \citep[e.g.][]{2009MNRAS.394.1085T, 2010MNRAS.401.1189F, 2015MNRAS.447..747L}.
\begin{equation}
    \label{Hamiltonian}
    \mathcal{H} = \mathcal{H}_{\rm quad} + \mathcal{H}_{\rm GR}
\end{equation}
with
\begin{equation}
    \mathcal{H}_{\rm quad} = \frac{1}{16} \left[\left(2 + 3e^{2} \right) \left(3 \cos^{2}i - 1\right) + 15e^{2}\sin^{2}i \cos2\omega \right] \ , 
\end{equation}
and
\begin{equation}
\mathcal{H}_{\rm GR} = \varepsilon_{\rm GR} \left(1 - e^2\right)^{-1/2} \ ,
\end{equation}
where
\begin{equation}
\varepsilon_{\rm GR} = \frac{3G \left(m_{\rm A} + m_{\rm B} \right)^{2}a_{\bullet}^3 \left(1 - e_{\bullet} \right)^{3/2}}{m_{\bullet} c^{2} a^{4} }\ ,
\end{equation}
and $c$ is the speed of light.

Lidov-Kozai oscillations can increase the binary's eccentricity to the point where it breaches its Roche limit. To quantify the maximum eccentricity, $e_{\rm max}$, reached by Lidov-Kozai oscillations, we impose conservation of energy and angular momentum to obtain \citep{2021MNRAS.502.2049L}
\begin{align}
    &\frac{3}{8}\Bigg\{j_{\rm{min}}^2-j_0^2+(5-4j_{\rm{min}}^2)
    \Bigg[1-\frac{\Big((j_{\rm{min}}^2-j_0^2)\eta-2j_0\cos i_0\Big)^2}{4j_{\rm{min}}^2}\Bigg]\nonumber
    \\
    &-(1+4e_0^2-5e_0^2\cos^2\omega_0)\sin^2i_0\Bigg\}+\varepsilon_{GR} \left(j_0^{-1}-j_{\rm{min}}^{-1}\right)=0 \,,
    \label{e_max}
\end{align}
where $e_0$, $\omega_0$, and $i_0$ are, respectively, the initial eccentricity, the initial argument of periapsis, and the initial mutual inclination of the binary orbit. 
We also define $j_{\rm min} = (1 - e_{\rm max}^2)^{1/2}$, $j_0 = (1 - e_0^2)^{1/2}$, and 
\begin{equation}
\eta = \frac{q}{(1+q)^{3/2}}  \sqrt{\frac{m_A a (1-e^2)}{m_{\bullet} a_{\bullet} (1-e_{\bullet}^2)} } \ .
\end{equation}
We then solve the implicit equation (\ref{e_max}) to obtain $e_{\rm max}$ for a given initial condition and check whether the binary's pericenter distance does not exceed its Roche limit, i.e., when
\begin{equation}
a \, (1 - e_{\rm max}) < r_{\rm R} \ .
\label{rochesurvive}
\end{equation}

To access the initial conditions affected by breakups through the Hill mechanism, we define the Hill radius following \cite{2019ApJ...875...42T}
\begin{equation}
    r_{\rm H} = \frac{a_{\bullet}}{2}(1 - e_{\bullet})\left(\frac{m_{\rm A} + m_{\rm B}}{3 m_{\bullet}}\right)^{1/3} \ .
\end{equation}
For a given initial condition, we then check whether the binary's apocenter distance does not exceed the Hill radius, i.e., when
\begin{equation}
a \, (1 + e_0) < r_{\rm H} \ . 
\label{hillsurvive}
\end{equation}

Given the exact same initial conditions from our numerical simulations, we finally select the surviving binaries from our sample, requiring that both conditions (\ref{rochesurvive}) and (\ref{hillsurvive}) are simultaneously satisfied.

In Fig.~\ref{Theo_vs_sim}, we compare the analytical predictions with the numerical results for the surviving systems, as well as for the systems that fall below the observational threshold.
We observe a remarkable agreement between the analytical predictions and the numerical simulations.
There is only a slight systematic overestimation by the analytical results that may be attributed to the simplifications that lead to the conditions (\ref{rochesurvive}) and (\ref{hillsurvive}), which do not account for higher order perturbations \citep[e.g.][]{2012MNRAS.424...52M, 2023MNRAS.522..937T, 2025arXiv250513780L}.

\section{Astrometric Signal}
\label{Appendix:AstrometricSignal}
In this section, we derive the equation for the projected astrometric signal.
We consider a frame $\left(X, Y, 0\right)$ attached to the orbit of the binary, with the $X-$axis is aligned with the pericenter.
In this frame, the orbit is an ellipse that can be expressed as
\begin{equation}
    \frac{X^{2}}{a^{2}} + \frac{Y^{2}}{b^{2}} = 1 \ , \label{eqellipse}
\end{equation}
where $a$ is the semi-major axis and $b = a \sqrt{1-e^2}$.
We now consider another frame $\left(x, y, z\right)$, where the $z-$axis is aligned with the line-of-sight, and thus the $xy-$plane corresponds to the plane of the sky.
The orbit of the binary system expressed in this new frame can be obtained through a succession of Euler rotations, $\mathcal{R}_{i}$, around the coordinate axes, such that
\begin{equation}
    \left(x,y,z \right)^{T} = \mathcal{R}_{3}\left(\Omega \right)\mathcal{R}_{1}\left(I \right)\mathcal{R}_{3}\left(\omega \right) \left(X,Y,0 \right)^{T} \ , 
\end{equation}
where $\Omega$ is the longitude of the ascending node, $I$ is the inclination to the line-of-sight, and $\omega$ is the argument of the pericenter.
The angle $\Omega$ only rotates the orientation of the projected ellipse in the plane of the sky and has no impact on the astrometric signal.
Therefore, for simplicity we set $\Omega = 0$ and obtain
\begin{align}
    &X = x  \cos \omega + y \sin \omega \sec I \label{X_equation} \ ,
    \\
    &Y = -x  \sin \omega + y \cos \omega \sec I \label{Y_equation} \ .
\end{align}
Inserting the previous expressions in equation (\ref{eqellipse}) gives a new generalized equation for the projected ellipse
\begin{equation}
    Ax^{2} + Bxy + Cy^{2} = 1 \ , 
\end{equation}
with
\begin{align*}
    &A = \frac{1 - e^{2}\cos^2 \omega}{a^{2} \left(1 - e^{2}\right)}\ ,
    \\
    &B = -\frac{e^{2} \sin 2 \omega \sec I}{a^{2}\left(1 - e^{2}\right)} \ ,
    \\
    &C = \frac{\left(1 - e^{2}\sin^2 \omega \right) \sec^2 I}{a^{2} \left(1 - e^{2}\right)} \ .
\end{align*}
The projected semi-major axis on the $xy-$plane is then
\begin{equation}
    a' = \sqrt 2 \, \left(\frac{A+C+\sqrt{\left(C-A\right)^{2}+B^{2}}}{4 A C-B^{2}}\right)^{1/2} \ ,
\end{equation}
or, simplifying,    
\begin{equation}
    a' = a \, \left(\gamma  + \sqrt{\gamma^2 - (1-e^2) \cos^2 I} \, \right)^{1/2}
    \label{semi_major} \ ,
\end{equation}
with
\begin{equation}
    \gamma = 1 - \frac{1}{2}\left[e^{2} 
 + \left(1-e^{2}\cos^2 \omega\right) \sin^2 I \, \right] \ .
\end{equation}

The astrometric semi-amplitude signal of a face-on elliptical orbit is given by \citep[e.g.][]{2017_Carroll}
\begin{equation}\label{astrometric_sin}
    \alpha = \frac{q}{1 + q}\frac{a}{D} = \frac{q}{\left(1 + q\right)^{2/3}}\frac{\left(G m_\mathrm{A} \right)^{1/3}}{n^{2/3} D} \ , 
\end{equation}
where $n=2 \pi / P$ is the mean motion, $a$ is the semi-major axis, $q$ is the mass ratio, and  $D$ is the distance from the observer to the celestial object. 
To account for an arbitrary orientation, we need to redefine the semi-major axis in  Eq.~(\ref{astrometric_sin}) using the projected semi-major axis given by Eq.~(\ref{semi_major}), resulting in
\begin{equation}\label{final_astrometric_signal}
    \alpha^{\prime} = \alpha \, \left(\gamma  + \sqrt{\gamma^2 - (1-e^2) \cos^2 I} \, \right)^{1/2} \ .
\end{equation}

\end{document}